\documentclass[reprint,twocolumn,aps,prl,amsmath,amssymb,floatfix,superscriptaddress,longbibliography]{revtex4-1}
\usepackage{graphicx}
 \usepackage{multirow}
\usepackage{epsfig}
\usepackage{bm}
\usepackage{dcolumn}
\usepackage{color}
\usepackage{physics}
\usepackage{float}
\usepackage[colorlinks,urlcolor=blue,citecolor=blue,linkcolor=magenta]{hyperref}
\makeatletter
\newcommand*{\rom}[1]{\expandafter\@slowromancap\romannumeral #1@}
\makeatother
\usepackage{tikz}
\usepackage{subfigure}

\begin{document}
	
\preprint{This line only printed with preprint option}

\title{Multifractal-enriched mobility edges and emergent quantum phases in Rydberg atomic arrays}

\author{Shan-Zhong Li}
\affiliation {Key Laboratory of Atomic and Subatomic Structure and Quantum Control (Ministry of Education), Guangdong Basic Research Center of Excellence for Structure and Fundamental Interactions of Matter, School of Physics, South China Normal University, Guangzhou 510006, China}
\affiliation {Guangdong Provincial Key Laboratory of Quantum Engineering and Quantum Materials, Guangdong-Hong Kong Joint Laboratory of Quantum Matter, Frontier Research Institute for Physics, South China Normal University, Guangzhou 510006, China}

\author{Yi-Cai Zhang}
\affiliation {School of Physics and Materials Science, Guangzhou University, Guangzhou 510006, China}

\author{Yucheng Wang}
\affiliation {Shenzhen Institute for Quantum Science and Engineering, Southern University of Science and Technology, Shenzhen 518055, China}

\author{Shanchao Zhang}
\affiliation {Key Laboratory of Atomic and Subatomic Structure and Quantum Control (Ministry of Education), Guangdong Basic Research Center of Excellence for Structure and Fundamental Interactions of Matter, School of Physics, South China Normal University, Guangzhou 510006, China}
\affiliation {Guangdong Provincial Key Laboratory of Quantum Engineering and Quantum Materials, Guangdong-Hong Kong Joint Laboratory of Quantum Matter, Frontier Research Institute for Physics, South China Normal University, Guangzhou 510006, China}
\affiliation{Quantum Science Center of Guangdong-Hong Kong-Macao Greater Bay Area, Shenzhen, China}

\author{Shi-Liang Zhu}
\email[Corresponding author: ]{slzhu@scnu.edu.cn}
\affiliation {Key Laboratory of Atomic and Subatomic Structure and Quantum Control (Ministry of Education), Guangdong Basic Research Center of Excellence for Structure and Fundamental Interactions of Matter, School of Physics, South China Normal University, Guangzhou 510006, China}
\affiliation {Guangdong Provincial Key Laboratory of Quantum Engineering and Quantum Materials, Guangdong-Hong Kong Joint Laboratory of Quantum Matter, Frontier Research Institute for Physics, South China Normal University, Guangzhou 510006, China}
\affiliation{Quantum Science Center of Guangdong-Hong Kong-Macao Greater Bay Area, Shenzhen, China}

\author{Zhi Li}
\email[Corresponding author: ]{lizphys@m.scnu.edu.cn}
\affiliation {Key Laboratory of Atomic and Subatomic Structure and Quantum Control (Ministry of Education), Guangdong Basic Research Center of Excellence for Structure and Fundamental Interactions of Matter, School of Physics, South China Normal University, Guangzhou 510006, China}
\affiliation {Guangdong Provincial Key Laboratory of Quantum Engineering and Quantum Materials, Guangdong-Hong Kong Joint Laboratory of Quantum Matter, Frontier Research Institute for Physics, South China Normal University, Guangzhou 510006, China}


\begin{abstract}
Anderson localization describes disorder-induced phase transitions, distinguishing between localized and extended states. In quasiperiodic systems, a third multifractal state emerges, characterized by unique energy and wave functions. {However, the corresponding multifractal-enriched mobility edges and three-state-coexisting quantum phases have yet to be experimentally detected. In this work, we propose exactly-solvable one-dimensional quasiperiodic lattice models that simultaneously host three-state-coexisting quantum phases, with their phase boundaries analytically derived via Avila’s global theorem.} Furthermore, we propose experimental protocols via Rydberg atom arrays to realize these states. { Notably, we demonstrate a spectroscopic technique capable of measuring inverse participation ratios  across real-space and dual-space domains, enabling simultaneous characterization of localized, extended, and multifractal quantum phases in systems with up to tens of qubits.} Our work opens new avenues for the experimental exploration of Anderson localization and multifractal states in artificial quantum systems.
\end{abstract}

\maketitle

{\emph{Introduction.}}-- Anderson localization elucidates a fundamental principle concerning how disorder induces a metal-insulator phase transition~\cite{PWAnderson1958}, which separates phases characterized by localized and extended states. Furthermore, three-dimensional periodic systems or even one-dimensional quasiperiodic lattices~\cite{SAubry1980} may exhibit phases where extended and localized states coexist, with a mobility edge (ME) distinguishing between them~\cite{NFMott1967}. Additionally, in quasiperiodic systems, a third fundamental state known as multifractal state emerges~\cite{FLiu2015,JWang2016}. The energy level statistics~\cite{TGeisel1991,SYJitomirskaya1999}, wave function distributions~\cite{TCHalsey1986,ADMirlin2006}, and dynamical properties~\cite{HHiramoto1988,RKetzmerick1997} of the multifractal state significantly differ from those of localized and extended states. The discovery of multifractal states has significantly broadened our understanding of Anderson localization. For instance, recent studies indicate that the multifractal state may enhance the superconducting transition temperature~\cite{JMayoh2015,MVFeigelman2007,MVFeigelman2010,ZFan2021,XZhang2022}.

Recent research has focused on identifying multifractal-enriched mobility edges (MMEs), which serve as boundaries separating multifractal states from either extended or localized states~\cite{XDeng2019,YWang2022a,YWang2022b,XLin2023,SZLi2023,RQi2023,QDai2023,CGuo2024,TLiu2022,YCZhang2022,ZWang2023,TLiu2024,XCZhou2023}. However, much of the progress in understanding MMEs has relied on numerical analyses, often involving tedious scaling assessments~\cite{YWang2022a,YWang2022b,XLin2023,SZLi2023,RQi2023,QDai2023,CGuo2024}. Recently, several approaches have been proposed to derive exact expressions for MMEs~\cite{TLiu2022,YCZhang2022,ZWang2023,TLiu2024,XCZhou2023,BSimon1989,MGoncalves2023,MGoncalves2023RG}. Despite significant efforts in studying MMEs~\cite{XDeng2019,YWang2022a,YWang2022b,XLin2023,SZLi2023,RQi2023,QDai2023,CGuo2024,TLiu2022,YCZhang2022,ZWang2023,TLiu2024,XCZhou2023,MGoncalves2023,MGoncalves2023RG,LZTang2022}, a fundamental question remains unanswered: Is there a universal platform capable of generating all types of MMEs and enabling the exploration of all possible multi-state coexisting quantum phases? Furthermore, while these states have been extensively investigated theoretically using LEs and IPRs~\cite{CGuo2024,TLiu2022,TLiu2024,YCZhang2022,XCZhou2023,MGoncalves2023RG,ZWang2023,YWang2020,YLiu2021,YWang2023LE,TLiu2024,LZTang2022,YLiu2021a,FEvers2008,JBiddle2010,JBiddle2011,SZLi2023a,DWZhang2020,HJiang2019,SZLi2024ATm,LWang2024a,LWang2024b,XCai2021,XCai2022,SGaneshan2015,SRoy2021,HYao2019}, these definitive indicators have yet to be experimentally observed. {This raises a critical question: How can we experimentally realize all MMEs and establish a practical protocol for their detection?}

In this Letter, we address these challenges by introducing a class of exactly solvable models that can be readily realized using Rydberg atom arrays. Specifically, we present a class of exactly solvable one-dimensional quasiperiodic flat band lattices, which host MMEs  and emergent quantum phases. All phase boundaries in these systems are analytically determined using Avila's global theorem~\cite{AAvila2015}, thereby circumventing the need for the tedious scaling analyses typically required in disordered systems~\cite{XDeng2019,YWang2022a,YWang2022b,XLin2023,SZLi2023,RQi2023,QDai2023,CGuo2024}.
Furthermore, we demonstrate that these models can be implemented in artificial quantum systems, such as superconducting quantum circuits and Rydberg atom arrays, and we provide a detailed realization scheme for Rydberg atom arrays. Remarkably, the key features of localized, extended, and multifractal states can be distinguished { in systems with up to tens of qubits} compared to several hundreds of qubits currently controllable in many research groups~\cite{SEbadi2021,CChen2023,Shaw2024}. {The critical problem in detecting MMEs lies in distinguishing extended states from multifractal regimes. While IPRs have been extensively employed in theoretical and numerical studies of Anderson localization and mobility edges ~\cite{TLiu2022,TLiu2024,YCZhang2022,XCZhou2023,MGoncalves2023,ZWang2023,YWang2020,YLiu2021,YWang2023LE,YLiu2021a,FEvers2008,JBiddle2010,JBiddle2011,SZLi2023a,DWZhang2020,HJiang2019,SZLi2024ATm,SZLi2023,LWang2024a,LWang2024b,XCai2021,XCai2022,SGaneshan2015,SRoy2021,HYao2019}, experimental observation of MMEs via IPR remains unexplored. Inspired by the method outlined in Ref. [31], we develop
a spectroscopic technique that simultaneously measures IPRs in real-space and dual-space domains, thereby resolving this problem.}

\begin{figure}[tbhp]	
\centering	
 \includegraphics[width=8.5cm]{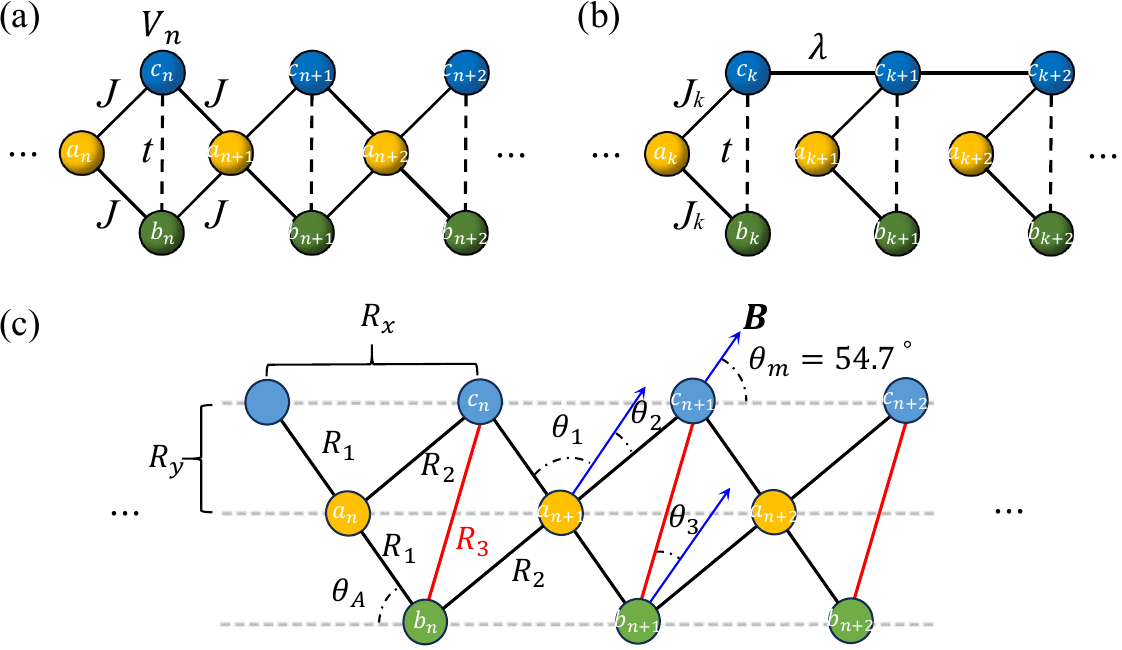}	
 \caption{The diamond lattice model represented in (a) lattice space, (b) dual space, (c) and its experimental implementation with Rydberg atomic array.}\label{F1}
\end{figure}

{\emph{The Diamond Lattice Model and Main Results.}}--We analytically demonstrate that MMEs can arise in a class of flat-band models featuring partially quasiperiodic modulation (see Supplementary Materials (SM)~\cite{Supplement}). As a representative example, we here utilize a diamond lattice~\cite{CDanieli2015} shown in  Fig.~\ref{F1}(a)   to illustrate our ideas, and the Hamiltonian of this model reads
\begin{equation}
\label{eqdim_main}
\begin{aligned}H_{D}=&\sum_{n=1}^{N}(Ja_{n}^{\dagger}b_{n}+Ja_{n}^{\dagger}c_{n}+tb_{n}^{\dagger}c_{n}+\mathrm{H.c.})\\&+\sum_{n=1}^{N-1}(Jb_{n}^{\dagger}a_{n+1}
+Jc_{n}^{\dagger}a_{n+1}+\mathrm{H.c.})+\sum_{n=1}^{N}V_{n}c_{n}^{\dagger}c_{n},
\end{aligned}
\end{equation}
where \(a_{n}\) (\(a_{n}^{\dagger}\)), \(b_{n}\) (\(b_{n}^{\dagger}\)), and \(c_{n}\) (\(c_{n}^{\dagger}\)) are the annihilation (creation) operators corresponding to sublattices \(A\), \(B\), and \(C\) in the \(n\)-th primitive cell, respectively. The quantities \(J\) and \(t\) denote the hopping strengths between the \(A\) and \(B/C\) sublattices and between the \(B\) and \(C\) sublattices, respectively. Here, \(N\) denotes the total number of primitive cells. A quasiperiodic potential, defined as \(V_{n}=2\lambda\cos(2\pi\alpha n+\theta)\), is applied solely to sublattice \(C\), where \(\lambda\), \(\alpha\), and \(\theta\) represent the strength of the quasiperiodic potential, an irrational number, and a phase offset, respectively. When \(\lambda=0\), Hamiltonian~\eqref{eqdim_main} showcases a perfect flat band characterized by \(E_{k}=-t\) along with two dispersive bands given by \(E_{k}=(t\pm\sqrt{16J^2\cos(k)+16J^2+t^2})/2\)~\cite{CDanieli2015}.

To accurately derive the MEs of Hamiltonian~\eqref{eqdim_main}, we employ a dual space as an auxiliary framework. By applying the dual transformations \(a_{n}=\frac{1}{\sqrt{N}}\sum_{k}a_{k}e^{-i2\pi\alpha kn}\), \(b_{n}=\frac{1}{\sqrt{N}}\sum_{k}b_{k}e^{-i2\pi\alpha kn}\), and \(c_{n}=\frac{1}{\sqrt{N}}\sum_{k}c_{k}e^{-i2\pi\alpha kn}\) for $\theta=0$, we can derive the corresponding Hamiltonian in dual space:
\begin{equation}\label{dimk}\begin{aligned}H_{K}=&\sum_{k=1}^{N}(J_{k}a_{k}^{\dagger}b_{k}+J_{k}a_{k}^{\dagger}c_{k}+\mathrm{H.c.})+\sum_{k=1}^{N}(t b_{k}^{\dagger}c_{k}+\mathrm{H.c.})\\&+\sum_{k=1}^{N-1}(\lambda c_{k+1}^{\dagger}c_{k}+\mathrm{H.c.}),\end{aligned}
\end{equation}
where \(J_{k}=J+Je^{(i2\pi\alpha k)}\). The geometric structure (refer to Fig.~\ref{F1}(b)) illustrates that the system operates as an extended Fano defect quasiperiodic lattice in dual space~\cite{CDanieli2015}. For the purpose of numerical calculations, we set \(J=1\) as the energy unit and impose periodic boundary conditions. The additional parameters are \(\theta=0\) and \(\alpha = \lim_{m\rightarrow\infty} \frac{F_{m-1}}{F_{m}}=(\sqrt{5}-1)/2\), where \(F_{m}\) denotes the \(m\)-th Fibonacci number. In finite-size studies, we specify the system size as \(N=F_{m}\) and \(\alpha=F_{m-1}/F_{m}\) to maintain accurate periodic boundary conditions. The MEs of Hamiltonian~\eqref{eqdim_main} can be categorized into two scenarios: \(t<2\) and \(t\ge2\). Given the similarity of outcomes in both cases, we present only the results for \(t<2\) in the main text (see SM~\cite{Supplement} for  the \(t\ge2\) case).

The primary findings of our analysis indicate that a comprehensive set of MMEs and all possible coexisting quantum phases can emerge within a class of flat-band models featuring partially quasiperiodic modulations. Furthermore, these predictions are readily demonstrable in current artificial quantum systems.

\begin{figure}[thbp]
	\centering	
    \includegraphics[width=8.5cm]{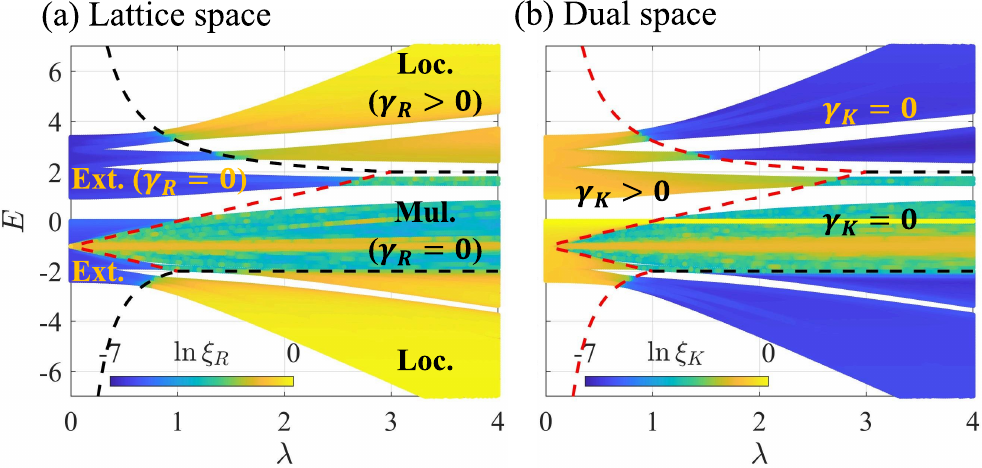}
	\caption{Phase diagram of the diamond lattice model. (a) The lattice space IPR \(\xi_{R}\) and (b) the dual space IPR \(\xi_{K}\) as functions of  potential strength \(\lambda\) and energy \(E\). The phase boundaries, marked by dashed lines, are determined from the critical energies that have been exactly solved. The parameters \(N=377\) and \(t=1\).
}\label{F2}
\end{figure}

The universal analytical expressions for the MMEs and the potential quantum phases of model~\eqref{eqdim_main} are consolidated in Table~\ref{tab1}. To facilitate comprehension of the universal expressions presented in this table, we depict results for a specific case in Fig.~\ref{F2}, which delineates three distinct regions. In the region \(\lambda \le 1\), we identify two types of MEs: one being a traditional ME that distinguishes between localized and extended states, and the other an MME that differentiates multifractal from extended states. In the region \(1 < \lambda < 3\), we observe three types of MEs: one separating localized and extended states, another MME separating multifractal and extended states, and the third MME distinguishing multifractal from localized states. Finally, in the region \(\lambda \ge 3\), only a single type of MME exists, effectively separating localized states from multifractal states, while the corresponding multi-state coexisting quantum phases also arise.

\begin{table*}[tbhp]
\renewcommand{\arraystretch}{2}
	\centering
 	\caption{MMEs and emergent quantum phases of model~\eqref{eqdim_main} under the condition of $t<2$}
	\begin{tabular}{|c |c| c| c| c| c| c|}
		\hline
		Disorder strength & \multicolumn{2}{c|}{$\lambda\le2-t$}                  & \multicolumn{3}{c|}{$2-t<\lambda<2+t$}                        & $\lambda\ge2+t$          \\  \hline

		Exact MMEs ($E_c=$)     & $\pm\frac{1}{\lambda}\pm\sqrt{\frac{1}{\lambda^2}\pm\frac{2t}{\lambda}+2}$ & $\pm\lambda-t$        &  $\frac{1}{\lambda}+\sqrt{\frac{1}{\lambda^2}+\frac{2t}{\lambda}+2}$   & $\lambda-t$       & -2                  & $\pm 2$         \\ \hline
		Separated states                 & Ext.$^*$ and Loc.                                     & Ext. and Mul.       & Ext. and Loc.                                                    & Ext. and Mul. & Ext. and Mul. & Loc. and Mul. \\ \hline
		Quantum phases                  & \multicolumn{2}{c|}{Ext.+Mul.;~Ext.+Mul.+Loc.}     & \multicolumn{3}{c|}{Ext.+Mul.+Loc.}                           & Loc.+Mul.            \\ \hline
	\end{tabular}	\label{tab1}
\vspace{0.5em}
$^*$Ext.=Extended states; \quad Loc.=Localized states; \quad  Mul.=Multifractal states.
\end{table*}

{\emph{Analytical expressions of the LEs.}}--- { In Anderson localization, the LE characterizes the inverse localization length of wavefunctions: A positive LE signifies exponential spatial decay (localized states), while a vanishing LE indicates extended states. The mobility edge, separating these phases, corresponds to critical LEs. } We now demonstrate that the LEs for the diamond lattice model can be analytically derived using Avila's global theory~\cite{AAvila2015}. The \(\gamma_{R(K)}\) defined as the LE of the eigenstate associated with the eigenvalue \(E\) in lattice (dual) space can be obtained from
\begin{equation}
\gamma_{R(K)}=\lim_{N\rightarrow\infty}\frac{1}{N}\ln\left \| {\textstyle \prod_{n(k)=1}^{N}}T_{n(k)} \right \|,
\end{equation}
where $\left \| \cdot \right\|$ denotes the matrix norm, and \(T_{n(k)}\) represents the transfer matrix.  The properties of the LEs are summarized in Table~\ref{tab2}~\cite{TLiu2024}.

The LE in lattice space is mathematically determined by the eigenequation of the Hamiltonian~\eqref{eqdim_main}: 
$\psi_{b,n-1}+\psi_{c,n-1}+\psi_{b,n}+\psi_{c,n}=E\psi_{a,n},$ $\psi_{a,n+1}+\psi_{a,n}+t\psi_{c,n}=E\psi_{b,n},$ $\psi_{a,n+1}+\psi_{a,n}+t\psi_{b,n}+V_{n}\psi_{c,n}=E\psi_{c,n}.$
By simplifying the equations, one can get
\begin{equation}
\begin{aligned}
\psi_{c,n+1}=&\frac{E^3-(E^2-2)V_{n}-t^2E-4(E+t)}{2(E+t)-V_{n+1}}\psi_{c,n}\\
&-\frac{2(E+t)-V_{n-1}}{2(E+t)-V_{n+1}}\psi_{c,n-1}.
\end{aligned}
\end{equation}
By extracting coefficients, one can obtain the transfer matrix as
$T_{n}=A_{n}B_{n}$, where  $A_{n}=1/M_{n+1}$ and
$$B_{n}=\left(\begin{matrix}
E^3-(E^2-2)V_{n}-t^2E-4(E+t)  & -M_{n-1} \\
M_{n+1} &0
\end{matrix}\right )
$$
with~$M_{n}=2(E+t)-V_{n}$.
Utilizing Avila's global theory~\cite{AAvila2015}, we derive analytical expressions for the LEs in terms of the eigenvalue \(E\) in lattice space. Similarly, we can also obtain the analytical LEs in dual space. The complete expressions for the LEs  can be written as

\begin{table}[tbhp]
\renewcommand{\arraystretch}{2}
	\centering
 	\caption{Key indicators of states' localization feature}
	\begin{tabular}{ c c c }
		\hline\hline
		States    &           LEs                                            &             IPRs                                     \\  \hline
		Ext.      &\quad $\gamma_{R}=0~\&~\gamma_{K}>0$            &\quad $\xi_{R}\sim 1/3N~\&~\xi_{K}(E)\sim\mathcal{O}(1)$        \\
		Loc.      &\quad $\gamma_{R}>0~\&~\gamma_{K}=0$            &\quad $\xi_{R}\sim\mathcal{O}(1)~\&~\xi_{K}\sim 1/3N$           \\
		Mul.      &\quad $\gamma_{R}=0~\&~\gamma_{K}=0$            &\quad $1/3N<\xi_{R}\approx \xi_{K}<\mathcal{O}(1)$              \\ \hline\hline
	\end{tabular}
	\label{tab2}
\end{table}

\begin{equation}\label{gammaRK}
(\gamma_{R},~\gamma_{K})=\left\{\begin{matrix}
(\gamma_{R,1},~\gamma_{K,1}),  &|E+t|> \lambda~\&~|E^2-2|> 2, \\
(\gamma_{R,2},~0), &|E+t|\le\lambda~\&~|E^2-2|> 2, \\
(0,~\gamma_{K2}), &|E+t|> \lambda~\&~|E^2-2|\le2,\\
(0,~0), &|E+t|\le\lambda ~\&~|E^2-2|\le2,
\end{matrix}\right.
\end{equation}
where $\gamma_{R,1}=\max\left\{\ln\left|\frac{\lambda c_{1}}{2c_{2}}\right|,0\right\}$, $\gamma_{R,2}=\ln\left|\frac{c_{1}}{2}\right|$, $\gamma_{K,1}=\max\left\{\ln\left|\frac{2c_{2}}{\lambda c_{1}}\right|,0\right\}$ and $\gamma_{K,2}=\ln\left|\frac{c_{1}}{\lambda}\right|$ with $c_{1}=|E^2-2|+\sqrt{(E^2-2)^2-4}$ and $c_{2}=|E+t|+\sqrt{(E+t)^2-\lambda^2}]$ ~\cite{Supplement}.

{\emph{Phase diagram determined by  the analytical LEs.}}---The MMEs and the emergent quantum phases listed in Table~\ref{tab1}, and the phase boundaries represented by dashed lines in Fig.~\ref{F2}, can be derived from the analytical expressions presented in Eq.~\eqref{gammaRK}. Mathematically, the inequality involving the absolute value yields two critical points. Consequently, each line in Eq.~\eqref{gammaRK} results in four critical points (\(E_c = -2, -t - \lambda, -t + \lambda, 2\)), which partition the energy axis into five distinct regions.


A specific value of \(\lambda\) results in three distinct relationships concerning the relative positions of the four critical points: \( -2 \leq -t - \lambda < -t + \lambda < 2 \), \( -t - \lambda < -2 < -t + \lambda < 2 \), and \( -t - \lambda < -2 < 2 \leq -t + \lambda\). Therefore, the discussion of the LEs must be divided into three cases: \textcircled{1}~\(\lambda \leq 2 - t\), \textcircled{2}~\(2 - t < \lambda < 2 + t\), and \textcircled{3}~\(\lambda \geq 2 + t\).

We can further derive the values of the LEs using the inequalities in Eq.~\eqref{gammaRK}. Case \textcircled{1} : In the regions where \(E < -2\) or \(E > 2\), we obtain the pairs \((\gamma_R, \gamma_K)=(\gamma_{R,1}, \gamma_{K,1})\) from the inequalities \(|E + t| > \lambda\) and \(|E^2 - 2| > 2\) presented in the first line of Eq.~\eqref{gammaRK}. Consequently, only extended states (\(\gamma_R = 0\) and \(\gamma_K > 0\)) or localized states (\(\gamma_R > 0\) and \(\gamma_K = 0\)) can exist within this energy interval. We can further derive two traditional MEs by setting \(\gamma_{R,1} = 0\): $E_c = \pm \frac{1}{\lambda} \pm \sqrt{\frac{1}{\lambda^2} \pm \frac{2t}{\lambda} + 2}. $ Moreover, the fourth line of Eq.~\eqref{gammaRK} indicates that \((\gamma_R, \gamma_K) = (0, 0)\) in the region \(-t - \lambda \leq E \leq -t + \lambda\), suggesting the emergence of multifractal states. Two corresponding MMEs can be identified: $ E_c = \pm \lambda - t.$ The remaining energy intervals can be analyzed similarly using the second and third lines of Eq.~\eqref{gammaRK}, leading to the conclusion that all eigenstates are extended. Thus, two-state (Extended + Multifractal) and three-state (Extended + Multifractal + Localized) coexisting quantum phases emerge under these conditions. Case \textcircled{2} : A similar examination can be conducted on expression \eqref{gammaRK}. The results indicate that while the ME \(E_c = \frac{1}{\lambda} + \sqrt{\frac{1}{\lambda^2} + \frac{2t}{\lambda} + 2}\) and the MME \(E_c = \lambda - t\) remain unchanged, the ME \(E_c = -\frac{1}{\lambda} - \sqrt{\frac{1}{\lambda^2} - \frac{2t}{\lambda} + 2}\) and the MME \(E = -\lambda - t\) merge into one point: \(E_c = -2\). Thus, only a three-state coexisting quantum phase emerges under these circumstances. Case \textcircled{3} : In this case, only two MMEs (\(E_c = \pm 2\)) are possible, leading to a two-state (Localized + Multifractal) coexisting quantum phase. This occurs because the ME \(E_c = \frac{1}{\lambda} + \sqrt{\frac{1}{\lambda^2} + \frac{2t}{\lambda} + 2}\) and the MME \(E_c = \lambda - t\) converge at \(E_c = 2\).

{\emph{Phases characterized with IPR.}}---We now analyze the IPR $\xi_{R(K)}$ of the eigenstates in lattice (or dual) spaces
\begin{equation}
\xi_{R(K)} = \sum_{n(k) = 1}^{N} \sum_{s = a, b, c} |\psi_{n(k), s}|^4,
\end{equation}
where \(\psi_{n(k), s}\) denote the wave functions on sublattice $s=\{ A, B,C\}$ within the \(n\) (\(k\))-th primitive cell. { The IPRs measure the spatial concentration of quantum states: a finite IPR (system-size-independent) signals localized states with wavefunctions peaked at few sites, while vanishing IPR ($\sim 1/N$) indicates extended states. The combined analysis of real-space and dual-space IPRs provides a practical method to distinguish multifractal, localized, and extended eigenstates. (see Table~\ref{tab2})}. As the quasiperiodic intensity \(\lambda\) increases, a region exhibiting multifractal characteristics emerges from the flat band at \(E = -t\). This observation supports the notion that multifractal states can arise from the flat band, ultimately leading to MMEs (see Fig.~\ref{F2}). It is noteworthy that the emergence of this multifractal region occurs exclusively when quasiperiodic modulation is applied to sublattices \(B\) or \(C\), both of which are associated with the flat band. Furthermore, similar MMEs can also manifest in diamond flat-band lattices, as well as in cross-stitch and Lieb flat-band lattices, when subjected to partial quasiperiodic modulation \cite{Supplement}.


 {\emph{Experimental realization on Rydberg atomic array.}}--- The lattice model in Eq.~\eqref{eqdim_main} can be realized in various artificial quantum systems, for specificity, we consider a Rydberg atomic array \cite{SEbadi2021,CChen2023,Shaw2024,Sdeleseleuc2019, VLienhard2020, ABrowaeys2020} to illustrate the experimental scheme. The Hamiltonian for the atomic array { with $N$ unit cells} in Fig.~\ref{F1}(c) can be expressed as
\begin{equation}\label{spinH_main}
\begin{aligned}
    H_{R}=&\sum_{n}(J_{AB}\sigma _{n,A}^{+}\sigma_{n,B}^{-}+J_{AC}\sigma _{n,A}^{+}\sigma_{n,C}^{-}+J_{BC}\sigma _{n,B}^{+}\sigma_{n,C}^{-}
\\&+J_{AB}\sigma _{n,B}^{+}\sigma_{n+1,A}^{-}+J_{AC}\sigma _{n,C}^{+}\sigma_{n+1,A}^{-}+\mathrm{H.c.} )\\
    &+\frac{1}{2}\sum_{n}V_{n}(1+\sigma_{n,C}^{z} ), 
\end{aligned}
\end{equation}
where \(\sigma^{\pm} = \frac{1}{2}(\sigma_x \pm i\sigma_y)\). The dipole-dipole interaction between Rydberg atoms is given by $J_{ij} = \frac{d^2}{R_{ij}^3}(3\cos^2\theta_{ij} - 1),$ where \(d\) represents the transition dipole moment between the two Rydberg levels, \(R_{ij}\) (with \(i, j = A, B, C\)) is the distance between sites \(i\) and \(j\), and \(\theta_{ij}\) is the angle between \(R_{ij}\) and the quantization axis defined by the magnetic field \(\mathbf{B}\)~\cite{Sdeleseleuc2019, VLienhard2020, ABrowaeys2020}. \(J_{ii}\) can be effectively mitigated to zero by selecting the magic angle \( \theta_{ii} = \theta_m = 54.7^{\circ}\).
In SM~\cite{Supplement}, we demonstrate that the model  in Eq.~\eqref{spinH_main} is equivalent to the model presented in Eq.~\eqref{eqdim_main} under the conditions that \( J_{AB} = J_{AC} = J \) and \( J_{BC} = t \).  To achieve identical coupling \( J_{ij} \) between sublattice \( A \) and sublattices \( B \) and \( C \), the following conditions must hold: $R_1 = \frac{2R_y}{\sin{\theta_A}}$,
$R_2 = \sqrt{R_1^2 + R_x^2 - 2R_1R_x\cos{\theta_A}},$ $R_3 = \sqrt{2R_1R_2\cos{(\theta_{1} + \theta_{2})}},$
where the angles $\theta_{1} = \pi - \theta_m - \theta_A,$ $ \theta_{2} = \theta_m - \arcsin{\left(\frac{R_1}{R_2}\sin{\theta_A}\right)},$ and $\theta_{3} = \theta_m - \arcsin{\left(\frac{2R_{1}}{R_3}\sin{\theta_A}\right)}.$
From these conditions, we can derive the coupling constants: $J = \frac{d^2}{R_1^3}(3\cos^2{\theta_1} - 1) = \frac{d^2}{R_2^3}(3\cos^2{\theta_2} - 1),$  and $t = \frac{d^2}{R_3^3}(3\cos^2{\theta_3} - 1)$.

The LEs and IPRs can be determined through the measurement of a quantity, denoted as { \(\text{P}_{\beta,m}^{R/K}\), which is defined subsequently. We follow the method outlined in Ref~\cite{PRoushan2017} to obtain \(\text{P}_{\beta,m}^{R}\).} The dynamics of the system governed by the Hamiltonian \(H_R\) satisfy the Schr\"{o}dinger equation: $|\psi(t)\rangle = e^{-iH_R t}|\psi(0)\rangle = \sum_{\beta} C_{\beta} e^{-iE_{\beta}t} |\psi_\beta\rangle$, where {\(C_{\beta}=\left\langle\psi_{\beta} | \psi(0)  \right \rangle \) and} \(\beta \in \{1, 2, 3, \ldots, 3N\}\) corresponds to the eigenvalue index. The initial state is selected as { $|\psi(0)\rangle_m = |0\rangle_1 \cdots |0\rangle_{m-1}  \left(\frac{|0\rangle_m + |1\rangle_m}{\sqrt{2}}\right) |0\rangle_{m+1}  \cdots |0\rangle_{3N}.$} During the evolution process, one can measure the time evolution curve of \(\langle \sigma^+_m \rangle = \langle \sigma^x_m \rangle + i \langle \sigma^y_m \rangle\), and subsequently apply a Fourier transform to obtain the squared modulus of the transformation for various frequencies \(\beta\), denoted as { \(\text{P}_{\beta,m}^R\), which represents the real space distribution in site $m$ for eigenvalue $E_\beta$ ~\cite{PRoushan2017,Supplement}. We can further do a Fourier transform to the real space distribution \(\text{P}_{\beta,m}^R\) to obtain the corresponding distribution \(\text{P}_{\beta,m}^K\) in the dual space.} Upon the derivation of \(\text{P}_{\beta,m}^{R/K}\), the corresponding IPR \(\xi_{^{R/K}}\) can be directly derived using the equation:
 {
\begin{equation}\label{expIPR}
\xi_{R/K}(E_{\beta}) = \sum_m (\text{P}_{\beta,m}^{R/K})^2.
\end{equation}
}
 Furthermore, the wave function of a localized state can be expressed as
\begin{equation}\label{fitting}
\psi_{R/K} \propto \max\{\text{P}_{\beta,m}^{R/K}\} e^{-\gamma_{R/K}(m-m_0)},
\end{equation}
where \(\max\{\text{P}^{R/K}_{\beta,m}\}\) represents the maximum amplitude at a fixed \(\beta\), and \(m_0\) corresponds to the site with the maximum amplitude.

\begin{figure}[htbp]
	\centering	
    \includegraphics[width=8cm]{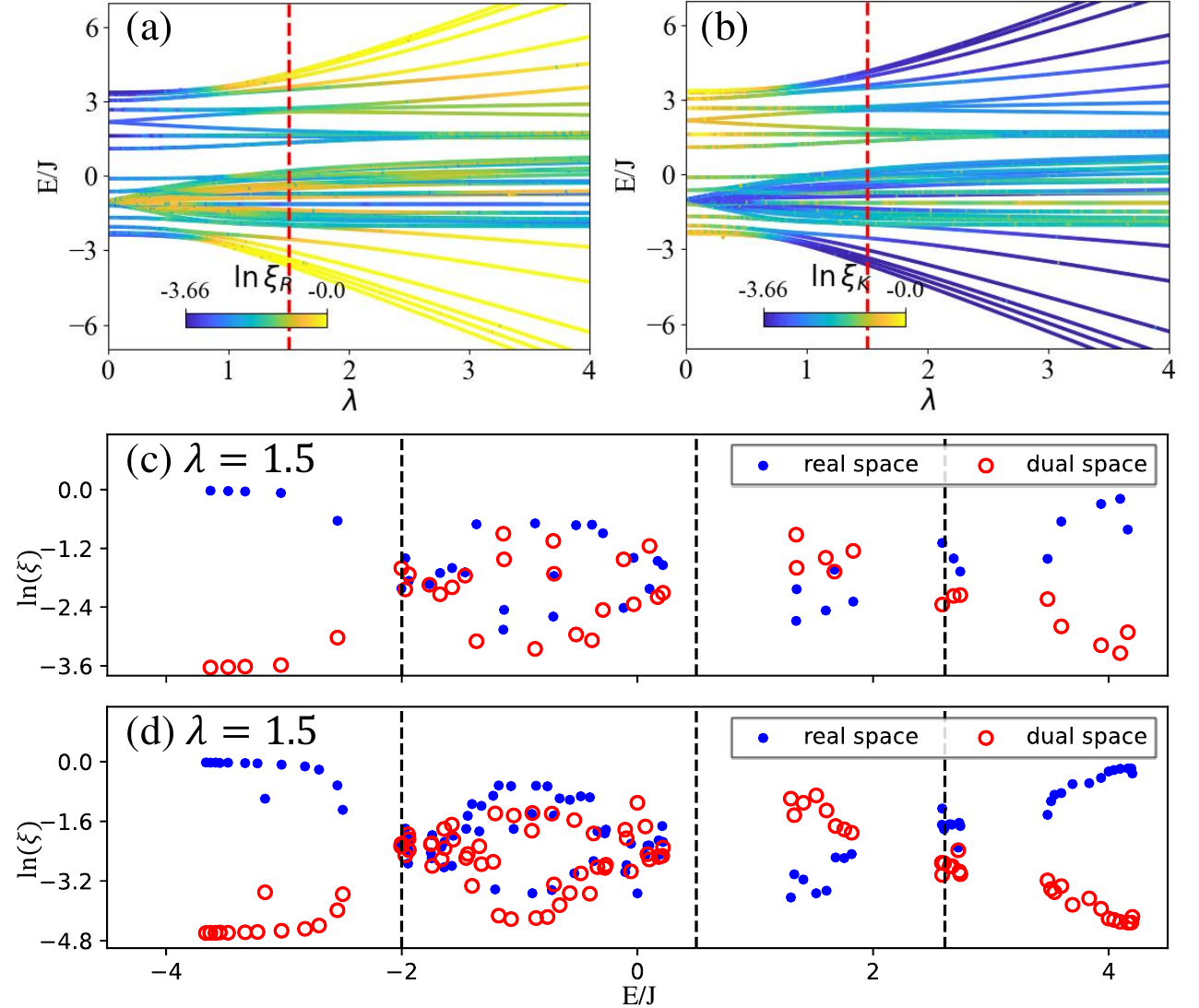}
	\caption{{The expected experimental IPRs. (a,b) The real and dual space IPRs \(\xi_{R/K}\)  as functions of  potential strength \(\lambda\) and energy \(E/J\) for $N=13$. Double space IPRs at \(\lambda=1.5\)  vs $E/J$ for (c) N=13 and (d) N=34.}}\label{Fig_IPRs}
\end{figure}

{ We have calculated the IPRs and LEs for systems with unit cell numbers selected based on Fibonacci sequences, specifically for system sizes N = 13, 21, 34, 55~\cite{Supplement}. Figure~\ref{Fig_IPRs} shows the experimentally measurable real-space and dual-space IPRs. Based on the analytical thresholds in Table~\ref{tab1}, we construct the phase diagram for $\lambda = 1.5$, revealing three distinct regimes: (i) localized states ($E < -2$ or $E > 2.61$), (ii) multifractal states ($-2 < E < 0.5$), and (iii) extended states ($0.5 < E < 2.61$). Remarkably, the experimentally measurable IPRs in Fig.~\ref{Fig_IPRs} closely align with the  criteria from Table~\ref{tab2}, enabling unambiguous phase discrimination. The key signatures are as follows: Localized/extended states: Real-space $\xi_R$ and dual-space $\xi_K$ IPRs exhibit spatial separation, with $\xi_R > \xi_K$ for localized states or $\xi_R < \xi_K$ for extended states. Multifractal states: $\xi_R$ and $\xi_K$ hybridize since $\xi_R \sim \xi_K$.
Notably, the results in Fig.~\ref{Fig_IPRs} demonstrate that 13-unit-cell systems suffice to capture the essential physics predicted by our model. This confirms that all three quantum phases (localized, extended, and multifractal states) are experimentally accessible with current Rydberg array platforms.}


{\emph{Conclusion.}}---We have introduced a class of one-dimensional, exactly solvable lattice models that exhibit a complete set of MMEs and multiple-state coexisting quantum phases.  Moreover, these models can be readily realized in artificial quantum systems, such as Rydberg atomic arrays and superconducting circuits. { Our results demonstrate that real and dual space IPRs, which are experimentally accessible, provide a criterion to unambiguously differentiate between these quantum phases.}

\emph{Acknowledgements.}---We thank Chang Li and Yan-Xiong Du for their insightful suggestions, Ling-Feng Yu and Rui-Jie Chen for their programming and computing support. This work was supported by the National Key Research and Development Program of China (Grant No.2022YFA1405300),  Innovation Program for Quantum Science and Technology (Grant No. 2021ZD0301700),  the Guangdong Basic and Applied Basic Research Foundation (Grant No.2021A1515012350), Guangdong Provincial Quantum Science Strategic Initiative(Grants No. GDZX2304002 and GDZX2404001), and the Open Fund of Key Laboratory of Atomic and Subatomic Structure and Quantum Control (Ministry of Education).

\global\long\def\id{\mathbbm{1}}
\global\long\def\ui{\mathbbm{i}}
\global\long\def\ud{\mathrm{d}}

\setcounter{equation}{0} \setcounter{figure}{0}
\setcounter{table}{0} 
\renewcommand{\theparagraph}{\bf}
\renewcommand{\thefigure}{S\arabic{figure}}
\renewcommand{\theequation}{S\arabic{equation}}

\onecolumngrid
\flushbottom
\newpage

{\begin{center}
		{\bf \large Supplementary materials for:\\Multifractal-enriched mobility edges and emergent quantum phases in Rydberg atomic arrays}
\end{center}}

\tableofcontents

\section{I. Quasiperiodic modulation on the "flat band irrelative" sublattice}
The Hamiltonian of the diamond structure lattice with quasiperiodic potential on the sublattice $A$ reads
\begin{equation}
H_{\text{R}}=\sum_{n=1}^{N}(Ja_{n}^{\dagger}b_{n}+Ja_{n}^{\dagger}c_{n}+tb_{n}^{\dagger}c_{n}+\mathrm{H.c.})+\sum_{n=1}^{N-1}(Jb_{n}^{\dagger}a_{n+1}+Jc_{n}^{\dagger}a_{n+1}+\mathrm{H.c.})+\sum_{n=1}^{N}V_{a,n}a_{n}^{\dagger}a_{n},
\end{equation}
and the corresponding eigenequations are
\begin{equation}
\begin{aligned}
&\psi_{b,n-1}+\psi_{c,n-1}+\psi_{b,n}+\psi_{c,n}+V_{n}\psi_{a,n}=E\psi_{a,n},\\
&\psi_{a,n+1}+\psi_{a,n}+t\psi_{c,n}=E\psi_{b,n},\\
&\psi_{a,n+1}+\psi_{a,n}+t\psi_{b,n}=E\psi_{c,n}.\\
\end{aligned}
\end{equation}
Comparing the equations in the second and third rows, one can get $\psi_{b,n}=\psi_{c,n}$. Substituting it into the equation in the first row, we obtain recursion formula about $\psi_{a}$
\begin{equation}
\psi_{a,n+1}+\frac{E-t}{2}V_{n}\psi_{a,n}+\psi_{a,n-1}=\frac{E^2-Et-4}{2}\psi_{a,n}.
\end{equation}

\begin{figure}[thbp]
\centering
\includegraphics[width=15cm]{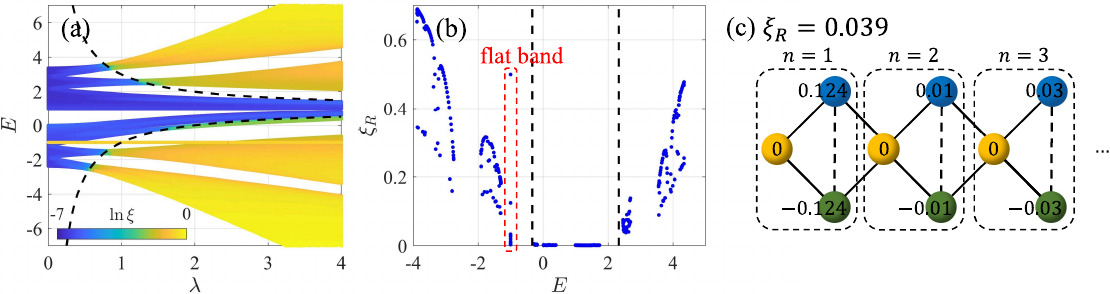}
\caption{(a) The IPR $\xi_{R}$ versus $\lambda$, where the black dashed line is the mobility edge (ME) $E_{c}$. (b) The IPR $\xi_{R}$ versus $E$ for $\lambda=1.5$, where the black dashed line is the ME $E_{c}=\pm\frac{4}{3}+1$. (c) Amplitudes of the flat-band $E=-t$ eigenstates with IPR $\xi_{R}= 0.039$ in the first three unit  cells at $\lambda = 1.5$. Throughout, $N = 377$ and $t = 1$.}\label{PA}
\end{figure}

Under the condition of $V=\frac{\lambda(E-t)}{2}$ and $E'=\frac{E^2-Et}{2}-2$, the result of the eigenequation is consistent with the traditional AA model, so the ME in the system can be determined by the following equation of the critical energy, i.e.,
\begin{equation}
\left|\frac{\lambda(E_{c}-t)}{2}\right|=1.
\end{equation}
Thus, we can obtain the MMEs' analytical expressions as
\begin{equation}
E_{c}=\pm\frac{2}{\lambda}+t.
\end{equation}
The results given by the analytic expression as well as that by the numerical IPR are both plotted in Fig.~\ref{PA}(a). The results reveal that, under such circumstances, there emerges no multifractal region in the system, which means the ME $E_{c}$ only separates the extended states from the localized states.

Besides, we find that the flat band at $E=-t$ is not destroyed as the quasiperiodic strength increases, which is significantly different from the case where the quasiperiodic modulation is exerted at the ``flat band related'' sublattice. As an example, we fix the parameter $\lambda = 1.5$, and show the corresponding IPR values of all eigenstates of the system with different $E$ in Fig.~\ref{PA}(b). The results demonstrate that there are still quite a number of eigenstates in the system at the flat band $E=-t$, which is the evidence that this quasiperiodic modulation cannot destroy the flat band.

Moreover, in Fig.~\ref{PA}(c), we show the amplitude distribution of the wave function for a specific eigenstate ($\xi_{R}= 0.039,~\lambda=1.5$) at the flat band energy. As in the case absent of quasiperiodic modulation ($\lambda=0$), all eigenstates corresponding to the flat band are the compact localized states~\cite{CDanieli2015}, i.e., states with the structure of $\psi_{n}=(0,1,-1)^{T}\delta_{n,n_{0}}/\sqrt{2}$, which will only appear in the $B$ and $C$ sublattice but not the $A$ sublattice. This again shows that MMEs can only be induced if quasiperiodic modulation is exerted on the "flat band related" sublattice.

\section{II. Derivation of Lyapunov exponents}
In the main text, we briefly introduce the LE in the lattice and dual spaces. Now, we exhibit a more detailed derivation process of LEs by means of Avila's global theory~\cite{AAvila2015}. The definition of the LE is
\begin{equation}
\gamma_{R(K)}=\lim_{N\rightarrow\infty}\frac{1}{N}\ln\left \| {\textstyle \prod_{n(k)=1}^{N}}T_{n(k)} \right \|,
\end{equation}
where $\left \|\cdot \right\|$ denotes the matrix norm. $T_{n(k)}$ represents the transfer matrix and $\gamma_{R(K)}$ denotes the LE of the eigenstate for the eigenvalue $E$ in lattice (dual) space. For localized (extended) eigenstates in lattice space, $\gamma_{R}>0$ and $\gamma_{K}=0$ ($\gamma_{R}=0$ and $\gamma_{K}>0$). For multifractal states, the wave function is delocalized in both spaces, so the LE satisfies $\gamma_{R}=\gamma_{K}=0$~\cite{TLiu2024}.

\subsection{II-1. Real space}
Here, we give a specific derivation of the real space LE. The Hamiltonian quantity of the real space is written as
\begin{equation}\label{eqdim}
H_{\text{R}}=\sum_{n=1}^{N}(Ja_{n}^{\dagger}b_{n}+Ja_{n}^{\dagger}c_{n}+tb_{n}^{\dagger}c_{n}+\mathrm{H.c.})+\sum_{n=1}^{N-1}(Jb_{n}^{\dagger}a_{n+1}+Jc_{n}^{\dagger}a_{n+1}+\mathrm{H.c.})+\sum_{n=1}^{N}V_{n}c_{n}^{\dagger}c_{n},
\end{equation}
and the corresponding eigenequation as
\begin{equation}
\psi_{c,n+1}=\frac{E^3-(E^2-2)V_{n}-t^2E-4(E+t)}{2(E+t)-V_{n+1}}\psi_{c,n}-\frac{2(E+t)-V_{n-1}}{2(E+t)-V_{n+1}}\psi_{c,n-1}.
\end{equation}
By extracting coefficients, one can obtain the corresponding transfer matrix as
\begin{equation}
T_{n}=A_{n}B_{n},
\end{equation}
where
\begin{equation}
\begin{aligned}
&A_{n}=1/M_{n+1}\\
&B_{n}=\left[\begin{matrix}
E^3-(E^2-2)V_{n}-t^2E-4(E+t)  & -M_{n-1} \\
M_{n+1} &0
\end{matrix}\right ]
\end{aligned}
\end{equation}
with~$M_{n}=2(E+t)-V_{n}$. According to the above expression, the LE can be divided into two parts, i.e., $\gamma_{R}=\gamma_{A}+\gamma_{B}$, in which~\cite{SLonghi2019a}
\begin{equation}\label{gammaA}
\gamma_{A}=\lim_{N\rightarrow\infty}\frac{1}{N}\ln\prod_{n=1}^{N}\frac{1}{\left|2(E+t)-2\lambda\cos(2\pi\alpha n+\theta)\right|}\\=\left\{\begin{matrix}
\ln\left|\frac{1}{|E+t|+\sqrt{(E+t)^2-\lambda^2}}\right|,  &|E+t|>\lambda, \\
\ln|\frac{1}{\lambda }|, &|E+t|\le\lambda.
\end{matrix}\right.
\end{equation}
For $\gamma_{B}$, we apply Avila's global theory of one-frequency analytical $SL(2,\mathbb{R})$ cocycle~\cite{AAvila2015}. The first step is to perform an analytical continuation of the global phase $\theta\rightarrow\theta+i\epsilon$ in $B_{n}$. In large $\epsilon$ limit, one can get
\begin{equation}
B_{n}=e^{-i2\pi\alpha n+\theta}e^{\epsilon}\begin{pmatrix}
-\lambda E^2-2\lambda & \lambda e^{i2\pi\alpha} \\
-\lambda e^{-i2\pi\alpha} &0
\end{pmatrix}+\mathcal{O}(1).
\end{equation}
According to Avila's global theory, $\gamma_{B}$, as a function of $\epsilon$, is a convex piecewise linear function with integer slopes~\cite{AAvila2015}. The discontinuity of the slope occurs when $E$ belongs to the spectrum of Hamiltonian~\eqref{eqdim} except for $\gamma_{B}=0$. Then, one can obtain
\begin{equation}\label{gammaB}
\gamma_B=\left\{\begin{matrix}
 \ln\left|\lambda\frac{|E^2-2|+\sqrt{(E^2-2)^2-4}}{2}\right|, & |E^2-2|> 2, \\
\ln|\lambda |,  & |E^2-2|\le 2.
\end{matrix}\right.
\end{equation}
By combining $\gamma_{A}$~\eqref{gammaA} and $\gamma_{B}$~\eqref{gammaB}, we obtain the LEs' analytical expressions with respect to the eigenvalue $E$ in lattice space. Similarly, we can also get LEs' expressions in dual space. The complete LEs' expression in different energy region can be written as
\begin{equation}\label{gammaR}
\gamma_{R}=\left\{\begin{matrix}
\max\left\{\ln\left|\dfrac{\lambda|E^2-2|+\lambda\sqrt{(E^2-2)^2-4}}{2|E+t|+2\sqrt{(E+t)^2-\lambda^2}}\right|,0\right\}, &|E+t|>\lambda~\&~|E^2-2|>2, \\
\max\left\{\ln\left|\dfrac{|E^2-2|+\sqrt{(E^2-2)^2-4}}{2}\right|,0\right\}, &|E+t|\le\lambda~\&~|E^2-2|>2, \\
\max\left\{\ln\left|\dfrac{\lambda}{|t+E|+\sqrt{(t+E)^2-\lambda}}\right|,0\right\}, &|E+t|>\lambda~\&~|E^2-2|\le2,\\
0, &|E+t|\le\lambda ~\&~|E^2-2|\le2.
\end{matrix}\right.
\end{equation}

Since the logarithmic function in the second (third) row of the expression~\eqref{gammaR} is always greater (smaller) than zero, $\gamma_{R}>0$ ($\gamma_{R}=0$) under the condition of $|E+t|\le\lambda~\&~|E^2-2|>2$ ($|E+t|>\lambda~\&~|E^2-2|\le2$). Then, one can simplify the LE's expression as
\begin{equation}\label{gammaRR}
\gamma_{R}=\left\{\begin{matrix}
\max\left\{\ln\left|\dfrac{\lambda|E^2-2|+\lambda\sqrt{(E^2-2)^2-4}}{2|E+t|+2\sqrt{(E+t)^2-\lambda^2}}\right|,0\right\}, &|E+t|>\lambda~\&~|E^2-2|>2, \\
\ln\left|\dfrac{|E^2-2|+\sqrt{(E^2-2)^2-4}}{2}\right|, &|E+t|\le\lambda~\&~|E^2-2|>2, \\
0, &|E+t|>\lambda~\&~|E^2-2|\le2,\\
0, &|E+t|\le\lambda ~\&~|E^2-2|\le2.
\end{matrix}\right.
\end{equation}
\subsection{II-2. Dual space}
The Hamiltonian in dual space reads
\begin{equation}\label{HK}
H_{K}=\sum_{k=1}^{N}(J_{k}a_{k}^{\dagger}b_{k}+J_{k}a_{k}^{\dagger}c_{k}+\mathrm{H.c.})+\sum_{n=1}^{N}(t b_{k}^{\dagger}c_{k}+\mathrm{H.c.})+\sum_{n=1}^{N-1}(\lambda c_{k+1}^{\dagger}c_{k}+\mathrm{H.c.}).
\end{equation}
Where $J_{k}=J+Je^{(i2\pi\alpha k)}$. Here, we insert a phase $\theta$ among $J_{k}$, i.e., $2\pi\alpha k\rightarrow(2\pi\alpha k+\theta)$, for the sake of the subsequent derivation of Avila's global theory. In fact, $\theta$ does not change the localization phase diagram, and in subsequent numerical calculations we set $\theta=0$.
From Hamiltonian~\eqref{HK}, one can obtain the corresponding eigenequation set, i.e.,
\begin{equation}
\begin{aligned}
& [1+e^{i(2\pi\alpha k+\theta)}]\psi_{b,k}+[1+e^{i(2\pi\alpha k+\theta)}]\psi_{c,k}=E\psi_{a,k},\\
&[1+e^{-i(2\pi\alpha k+\theta)}]\psi_{a,k}+t\psi_{c,k}=E\psi_{b,k},\\
&\lambda\psi_{c,k+1}+\lambda\psi_{c,k-1}+[1+e^{-i(2\pi\alpha k+\theta)}]\psi_{a,k}+t\psi_{b,k}=E\psi_{c,k}.\\
\end{aligned}
\end{equation}
By combining the first and second rows of the above eigenequation set, we have
\begin{equation}\label{dualspace1}
\psi_{a,k}=\frac{(t+E)[1+e^{i(2\pi\alpha k+\theta)}]}{E^2-2-2\cos(2\pi\alpha k+\theta)}\psi_{c,k},
\end{equation}
and
\begin{equation}\label{dualspace2}
\psi_{b,k}=\frac{tE+2+2\cos(2\pi\alpha k+\theta)}{E^2-2-2\cos(2\pi\alpha k+\theta)}\psi_{c,k}.
\end{equation}
Then, one can obtain new eigenequations for the component $\psi_{c}$, i.e.,
\begin{equation}
\psi_{c,k+1}=-\frac{4t+4E+t^2E-E^3-4(t+E)\cos(2\pi\alpha k+\theta)}{\lambda[E^2-2-2\cos(2\pi\alpha k+\theta)]}\psi_{c,k}-\psi_{c,k-1}.
\end{equation}
Through the above equation, one can calculate the corresponding transfer matrix as
\begin{equation}
\begin{aligned}
T_{k}=A_{k}B_{k},
\end{aligned}
\end{equation}
where
\begin{equation}
\begin{aligned}
&A_{k}=\frac{1}{\lambda[E^2-2-2\cos(2\pi\alpha k+\theta)]},\\
&B_{k}=\left[\begin{matrix}
-4t-4E-t^2E+E^3+4(t+E)\cos(2\pi\alpha k+\theta) & -\lambda[E^2-2-2\cos(2\pi\alpha k+\theta)]\\
\lambda[E^2-2-2\cos(2\pi\alpha k+\theta)] &0
\end{matrix}\right ].
\end{aligned}
\end{equation}
The LE can be computed by $\gamma_{K}(E)=\gamma_{A}(E)+\gamma_{B}(E)$, in which~\cite{SLonghi2019a}
\begin{equation}
\begin{aligned}
\gamma_{A}&=\lim_{N\rightarrow\infty}\frac{1}{N}\ln\prod_{k=1}^{N}\frac{1}{\lambda[E^2-2-2\cos(2\pi\alpha k+\theta)]}=\frac{1}{2\pi}\int_{0}^{2\pi}\ln{\frac{1}{\left|\lambda[E^2-2-2\cos(\phi)]\right|}}d\phi \\
&=\left\{\begin{matrix}
\ln\left|\dfrac{2}{|\lambda E^2-2\lambda |+\sqrt{(\lambda E^2-2\lambda )^2-4\lambda^2}}\right|, &|E^2-2|>2, \\
\ln|\dfrac{1}{\lambda }|, &|E^2-2|\le2.
\end{matrix}\right.
\end{aligned}
\end{equation}

As for $\gamma_{B}$, one can use the Avila's global theory. The first step is to perform an analytical continuation of the global phase $\theta\rightarrow\theta+i\epsilon$ in $B_{k}$. In large $\epsilon$ limit, one can get
\begin{equation}
B_{k}=e^{-i2\pi\alpha k+\theta}e^{\epsilon}\begin{pmatrix}
2(t+E) & \lambda \\
-\lambda &0
\end{pmatrix}+\mathcal{O}(1).
\end{equation}
According to Avila's global theory, $\gamma_{B}$, as a function of $\epsilon$, is a convex piecewise linear function with integer slopes~\cite{AAvila2015}. One can obtain
\begin{equation}
\gamma_B=\left\{\begin{matrix}
\ln\left|2|t+E|+2\sqrt{(t+E)^2-\lambda^2}\right|, & |E+t|>\lambda, \\
\ln|\lambda|, & |E+t|\le\lambda.
\end{matrix}\right.
\end{equation}
By combining the information of $\gamma_{A}$ and $\gamma_{B}$, one can obtain the corresponding LE versus $E$ as
\begin{equation}\label{gammaK}
\gamma_{K}=\left\{\begin{matrix}
\max\left\{\ln\left|\dfrac{2|E+t|+2\sqrt{(E+t)^2-\lambda^2}}{\lambda|E^2-2|+\lambda\sqrt{(E^2-2)^2-4}}\right|,0\right\}, &|E+t|>\lambda~\&~|E^2-2|>2, \\
\max\left\{\ln\left|\dfrac{2}{|E^2-2|+\sqrt{(E^2-2)^2-4}}\right|,0\right\}, &|E+t|\le\lambda~\&~|E^2-2|>2, \\
\max\left\{\ln\left|\dfrac{|t+E|+\sqrt{(t+E)^2-\lambda}}{\lambda}\right|,0\right\}, &|E+t|>\lambda~\&~|E^2-2|\le2,\\
0, &|E+t|\le\lambda ~\&~|E^2-2|\le2.
\end{matrix}\right.
\end{equation}
Since the logarithmic function in the second row of the expression~\eqref{gammaK} is always less than zero, $\gamma_{K}=0$ under the condition of $|E+t|\le\lambda~\&~|E^2-2|>2$. Then, one can simplify the LE's expression as
\begin{equation}\label{gammaKK}
\gamma_{K}=\left\{\begin{matrix}
\max\left\{\ln\left|\dfrac{2|E+t|+2\sqrt{(E+t)^2-\lambda^2}}{\lambda|E^2-2|+\lambda\sqrt{(E^2-2)^2-4}}\right|,0\right\}, &|E+t|>\lambda~\&~|E^2-2|>2, \\
0, &|E+t|\le\lambda~\&~|E^2-2|>2, \\
\ln\left|\dfrac{|t+E|+\sqrt{(t+E)^2-\lambda}}{\lambda}\right|, &|E+t|>\lambda~\&~|E^2-2|\le2,\\
0, &|E+t|\le\lambda ~\&~|E^2-2|\le2.
\end{matrix}\right.
\end{equation}
The traditional ME is given by $\gamma_R=\gamma_K=0$ for the first line of the Eq.~\eqref{gammaRR} and \eqref{gammaKK}, thus one can obtain four critical points on the energy axis, i.e.,
\begin{equation}
E_{c}=\left\{\begin{matrix}
\dfrac{1}{\lambda} +\sqrt{\dfrac{1}{\lambda^2}+\dfrac{2t}{\lambda}+2 }, \\
\dfrac{1}{\lambda} -\sqrt{\dfrac{1}{\lambda^2}+\dfrac{2t}{\lambda}+2 }, \\
- \dfrac{1}{\lambda} +\sqrt{\dfrac{1}{\lambda^2}-\dfrac{2t}{\lambda}+2 },\\
-\dfrac{1}{\lambda} -\sqrt{\dfrac{1}{\lambda^2}-\dfrac{2t}{\lambda}+2 }.
\end{matrix}\right.
\end{equation}
Note that, $E_{c}$ can only emerge in the energy region $|E+t|>\lambda~\&~|E^2-2|>2$. Therefore, only two of the four critical points can be chosen, i.e., $E_{c,1}=\frac{1}{\lambda} +\sqrt{\frac{1}{\lambda^2}+\frac{2t}{\lambda}+2}$ and $E_{c,2}=-\frac{1}{\lambda} -\sqrt{\frac{1}{\lambda^2}-\frac{2t}{\lambda}+2 }$. By considering the values of different parameters $\lambda$, one can directly obtain LEs in different cases.

\section{III. From analytical expressions to phase diagram}

For diamond lattice in the main text, the complete phase diagram consists of two parts, namely, the condition of $t<2$ and $t\ge2$. In this section, we will detail the complete process of obtaining phase diagram from analytic expression.

\subsection{III-1. The case of $t<2$}
First, we discuss the case of $t<2$. Now, we know that the LEs analytic expressions of lattice space and dual space are, respectively,
\begin{equation}\label{GGR}
\gamma_{R}=\left\{\begin{matrix}
\gamma_{R,1}, &|E+t|> \lambda~\&~|E^2-2|> 2, \\
\gamma_{R,2}, &|E+t|\le\lambda~\&~|E^2-2|> 2, \\
0, &|E+t|> \lambda~\&~|E^2-2|\le2,\\
0, &|E+t|\le\lambda ~\&~|E^2-2|\le2,
\end{matrix}\right.
\end{equation}
where $\gamma_{R,1}=\max\left\{\ln\left|\lambda\dfrac{|E^2-2|+\sqrt{(E^2-2)^2-4}}{2|E+t|+2\sqrt{(E+t)^2-\lambda^2}}\right|,0\right\}$, $\gamma_{R,2}=\ln\left|\dfrac{|E^2-2|+\sqrt{(E^2-2)^2-4}}{2}\right|$,
and
\begin{equation}\label{GGK}
\gamma_{K}=\left\{\begin{matrix}
\gamma_{K,1}, &|E+t|>\lambda~\&~|E^2-2|>2, \\
0, &|E+t|\le\lambda~\&~|E^2-2|>2, \\
\gamma_{K,2}, &|E+t|>\lambda~\&~|E^2-2|\le2,\\
0, &|E+t|\le\lambda ~\&~|E^2-2|\le2,
\end{matrix}\right.
\end{equation}
where $\gamma_{K,1}=\max\left\{\ln\left|\dfrac{2|E+t|+2\sqrt{(E+t)^2-\lambda^2}}{\lambda|E^2-2|+\lambda\sqrt{(E^2-2)^2-4}}\right|,0\right\}$, $\gamma_{K,2}=\ln\left|\dfrac{|t+E|+\sqrt{(t+E)^2-\lambda}}{\lambda}\right|$.

\begin{figure}[thbp]
\centering
\includegraphics[width=15cm]{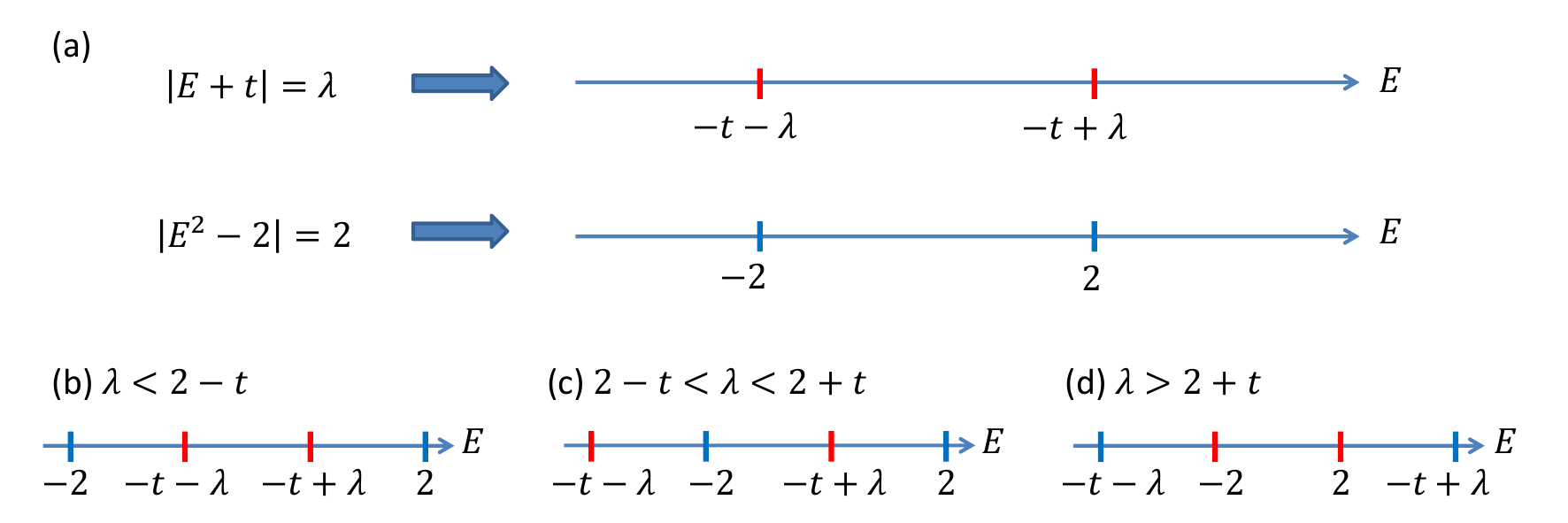}
\caption{(a) Four critical points on the energy axis. (b-d) Three possible relative positions of the four critical points.}\label{FS2a}
\end{figure}

From the equations~\eqref{GGR} and~\eqref{GGK}, one can see that the values of LEs are determined by two inequalities no matter in the lattice space or in the dual space. Mathematically, since an inequality with an absolute value operation has two critical points, two inequalities will give four critical points [see Fig.~\ref{FS2a}]. Obviously, there are three different relationships of relative position between the four critical points, i.e.,

\begin{equation}
\begin{aligned}
&\textcircled{1} -2<-t-\lambda<-t+\lambda<2~~~~\text{for}~\lambda\le 2-t,\\
&\textcircled{2} -t-\lambda<-2<-t+\lambda<2~~~~\text{for}~2-t\le\lambda\le 2+t,\\
&\textcircled{3} -t-\lambda<-2<2<-t+\lambda~~~~\text{for}~\lambda\ge 2+t.\\
\end{aligned}
\end{equation}

\begin{figure}[thbp]
\centering
\includegraphics[width=15cm]{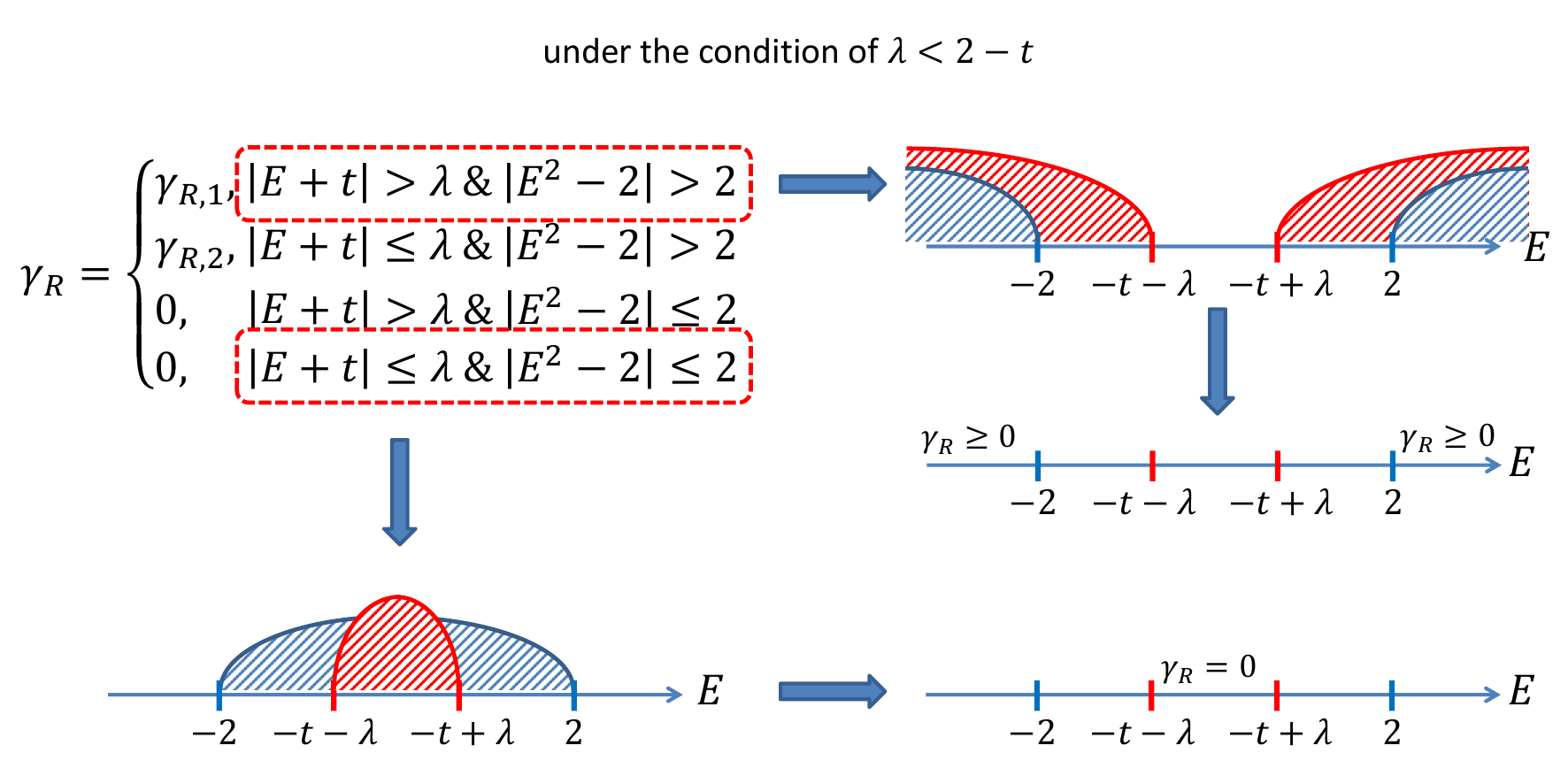}
\caption{The LEs for different regions determined by the analytical expressions.}\label{FS2b}
\end{figure}

Furthermore, after determining the relative positions of the four critical points, one can obtain the values of LEs in different ranges on the energy axis through the information given by the inequalities. For example, let's consider case \textcircled{1}. We plot the process of getting the LEs' value on the energy axis in Fig.~\ref{FS2b}. As shown in the figure, we obtain that in the region of $E<-2$ or $E>2$, the LE $\gamma_R=\gamma_{R,1}$. In other words, both $\gamma_{R}=\ln\left|\lambda\dfrac{|E^2-2|+\sqrt{(E^2-2)^2-4}}{2|E+t|+2\sqrt{(E+t)^2-\lambda^2}}\right|$ and $\gamma_R=0$ are valid in this region. Then, we obtain $\gamma_{R}\ge0$ in the region of $E<-2$ and $E>2$ [see Fig.~\ref{FS2b}]. Perform the same analysis on the fourth line of the expression, we obtain $\gamma=0$ in the region of $-t-\lambda< E< -t+\lambda$.

By analyzing the inequality information given by each line of the expression~\eqref{GGR} and~\eqref{GGK} step by step, one can obtain all LEs, which includes results both in the lattice and the dual spaces. By combining the above information from the two dual spaces, one can obtain the complete set of MMEs and all possible emergent quantum phases.

\subsection{III-2. The case of $t\ge2$}

Now, we turn to the case of $t\ge2$. The main analysis process is the same as previous subsection III-1. Under such circumstances, the critical points generated by the inequality also have three relative positions on the energy axis $E$, i.e.,
\begin{equation}
\begin{aligned}
&\textcircled{1}~t-\lambda<-t+\lambda<-2<2~~~~for~\lambda\le t-2,\\
&\textcircled{2}~t-\lambda<-2<-t+\lambda<2~~~~for~t-2\le\lambda\le t+2,\\
&\textcircled{3}~t-\lambda<-2<2<-t+\lambda~~~~for~\lambda\ge t+2.\\
\end{aligned}
\end{equation}

\begin{figure}[thbp]
\centering
\includegraphics[width=15cm]{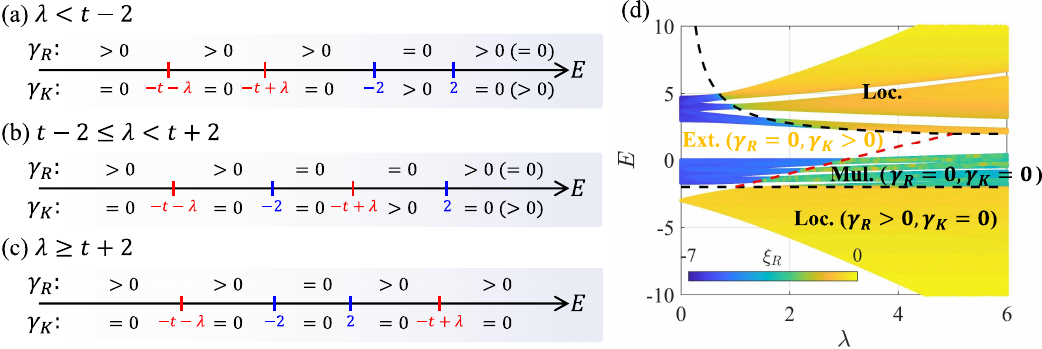}
\caption{LEs' Phase diagram versus $E$ with different $\lambda$. (b) The lattice space IPR $\xi_{R}$ versus $\lambda$, where the black (red) dashed line is the critical energy separating $\gamma_{R}>0$ and $\gamma_{R}=0$ regions ($\gamma_{K}>0$ and $\gamma_{K}=0$ regions) in lattice (dual) space. Throughout, $N=377$ and $t=1$.}\label{t3}
\end{figure}

The complete phase diagram can be obtained using the same analysis method as in the previous Sec.~II-1. We plot the phase diagram on energy axis $E$ for different $\lambda$ in Fig.~\ref{t3}(a-c). The results comfirm again that all types of MMEs and multi-state coexisting quantum states emerge.

Besides, the numerically calculated IPR is also shown in Fig.~\ref{t3} to indicate the emergence of MMEs (in the region of $\lambda>t-2$), which agrees perfectly with the theoretical expression (dashed lines). We summarize the results corresponding to the case of $t\ge 2$ in Tab.~\ref{tabt3}.

\begin{table*}[tbhp]
\renewcommand{\arraystretch}{2}
\centering
\caption{MMEs and emergent quantum phases of flat-band partially-quasiperiodic diamond lattice for the case of $t\ge2$.}
\begin{tabular}{|c |c| c |c| c| c| c|}
\hline
Quasiperiodic strength & \multicolumn{2}{c|}{$\lambda< t-2$} & \multicolumn{3}{c|}{$t-2\le\lambda<2+t$} & $\lambda\ge t+2$ \\ \hline
Exact MMEs ($E_c=$) & $\frac{1}{\lambda}+\sqrt{\frac{1}{\lambda^2}+\frac{2t}{\lambda}+2}$ & -2 & $\frac{1}{\lambda}+\sqrt{\frac{1}{\lambda^2}+\frac{2t}{\lambda}+2}$ & $\lambda-t$ & -2 & $\pm 2$ \\ \hline
Separated states & Ext.$^*$ and Loc. & Ext. and Loc. & Ext. and Loc. & Ext. and Mul. & Ext. and Mul. & Loc. and Mul. \\ \hline
Possible phases & \multicolumn{2}{c|}{Ext.+Loc.} & \multicolumn{3}{c|}{Ext.+Mul.+Loc.} & Loc.+Mul. \\ \hline
\end{tabular}
\label{tabt3}
\vspace{0.5em}
$^*$Ext.=Extended states; \quad Loc.=Localized states; \quad Mul.=Multifractal states.
\end{table*}

\section{IV. The corresponding critical exponents}
\begin{figure}[bthp]
\centering
\includegraphics[width=8.5cm]{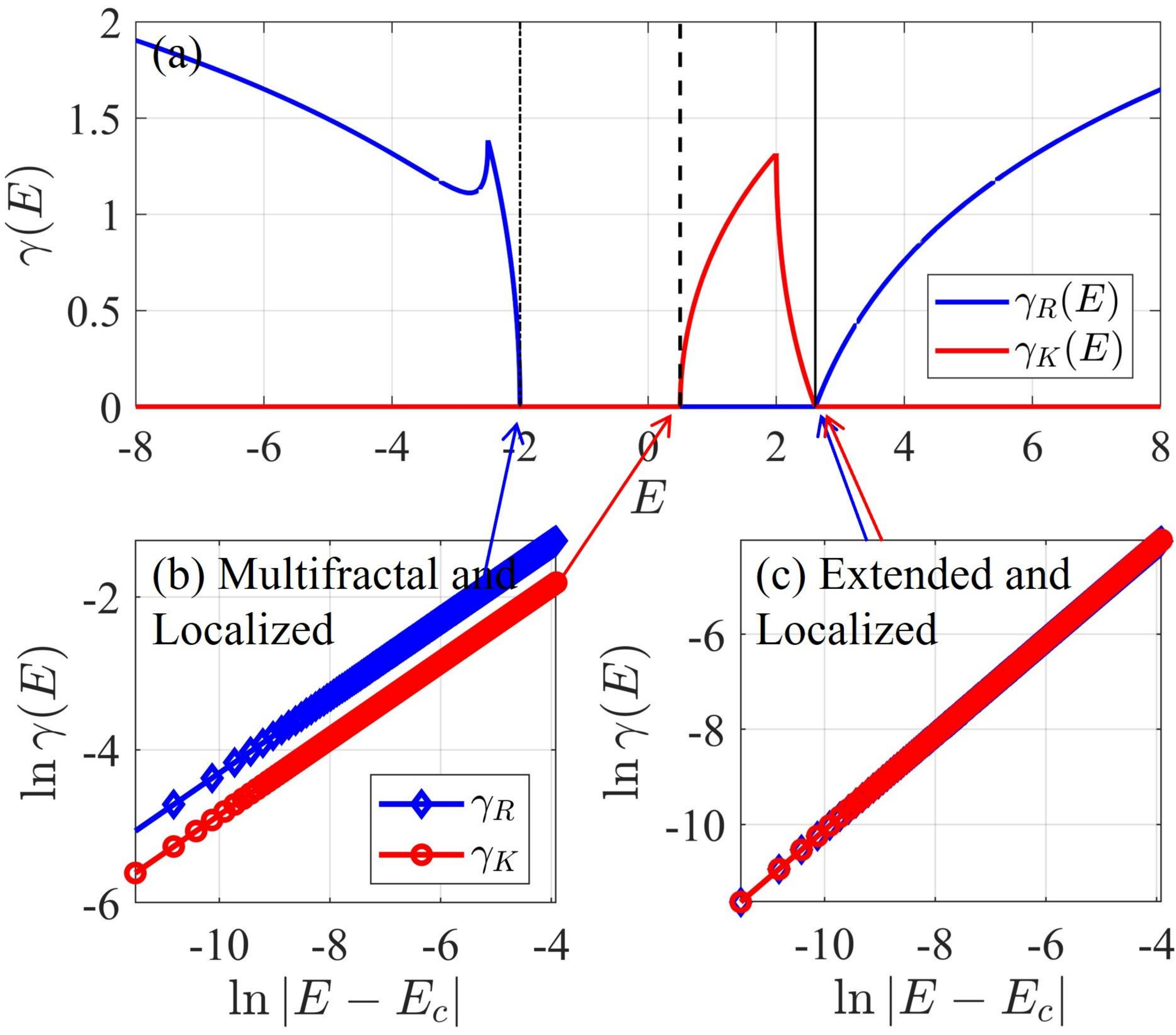}
\caption{(a) The LE versus $E$ in the region of $E\in\left[-8,8\right]$. $\ln\gamma$ versus $\ln|E-E_c|$ for the case of $E_c= -2$ and $E_c=0.5$ (b), and the case of $E = 2.6103$ (c). The red and blue lines stand for the LEs in the lattice and dual spaces, respectively. Other parameters $t=1$ and $\lambda=1.5$. }\label{S5}
\end{figure}
Critical exponent, as an important indicator of the universal class of phase transitions, has been widely used in the study of localized phase transitions. The standard Anderson transition from the extended phase to the localized phase corresponds to a critical exponent $\mu=1$~\cite{FEvers2008}, while the corresponding critical exponent of the transition from the multifractal phase to the localized phase is 0.5~\cite{YCZhang2022}.

Here we calculate the critical exponent of the model in the main text and plot that in Fig.~\ref{S5}.

First, we exhibit the LE in lattice and dual space for all eigenenergies between $E\in\left[-8,8\right]$ at $\lambda = 1.5$ and $t=1$ [see Fig.~\ref{S5}(a)]. It can be seen clearly that the LEs for ME separating the multifractal and localized states are very different from the LE for ME separating the extended and localized states. By fitting the LEs under log-log scale, we obtain the corresponding crtical exponents, which equal to the slope of $\gamma_{R(K)}$ under log-log scale in Fig.~\ref{S5}(b)(c). On the one hand, from Fig.~\ref{S5}(b), one can find that the critical exponent $\nu=1/2$ for the critical energy separating the multifractal and localized states ($E_c=-2$ for $\gamma_R$ and $E_c=0.5$ for $\gamma_K$). On the other hand, from Fig.~\ref{S5}(c), the correpsonding critical exponent $\nu=1$ for the critical energy separating the extended and localized states ($E_c=2.6103$ for both $\gamma_R$ and $\gamma_K$). The critical exponent again from another perspective supports the correctness of the results given by analyzing MMEs and IPR in the main text.

\section{V. More evidence to support the universality of the theory}

In order to prove the universality of the theory, we provide another two typical flat-band partially-quasiperiodic lattice models, namely, quasiperiodic cross-stitch lattice and quasiperiodic Lieb lattice.

\subsection{V-1. Quasiperiodic cross-stitch lattice}

First, we discuss MMEs and the emergent multi-state coexisting quantum phase in quasiperiodic cross-stitch lattice.

\begin{figure}[thbp]
\centering
\includegraphics[width=8cm]{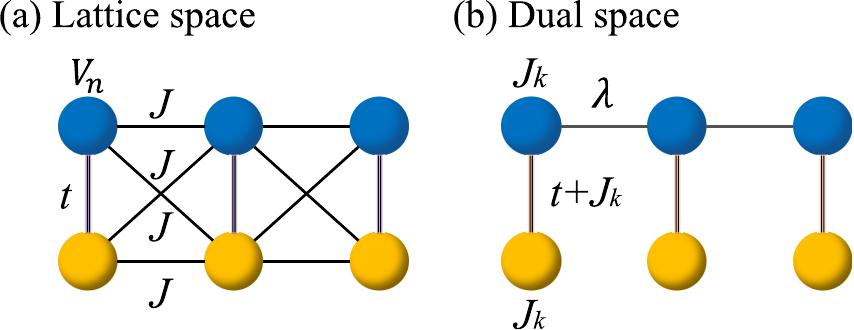}
\caption{The schematic diagram of the quasiperiodic cross-stitch lattice in lattice space (a) and in dual space (b). Blue and yellow balls correspond to sublattice A and B, respectively.}\label{figcross}
\end{figure}

The geometric structure is shown schematically in Fig.~\ref{figcross}(a) and the Hamiltonian can be written as
\begin{equation}\label{RealCS}
H_{R}=\sum_{n=1}^{N-1}J(a_{n}^{\dagger}b_{n+1}+a_{n}^{\dagger}a_{n+1}+b_{n}^{\dagger}a_{n+1}+b_{n}^{\dagger}b_{n+1}+\mathrm{H.c.})+\sum_{n=1}^{N}(ta_{n}^{\dagger}b_{n}+\mathrm{H.c.})+\sum_{n=1}^{N}V_{n}a_{n}^{\dagger}a_{n},
\end{equation}
where $a_{n}$ ($a_{n}^{\dagger}$) and $b_{n}$ ($b_{n}^{\dagger}$) represent the annihilation (creation) operators of sublattices $A$ and $B$ in the $n$-th unit cell, respectively. $J$ and $t$ denote the inter- and intra-hopping strength, which are marked in the Fig.~\ref{figcross}. $N$ represents the total number of unit  cells. The quasiperiodic potential $V_{n}=2\lambda\cos(2\pi\alpha n+\theta)$ is applied only on the sublattice $A$, where $\lambda$, $\alpha$, and $\theta$ denote the quasiperiodic strength, an irrational number, and a phase offset, respectively. Under the condition of $\lambda=0$, Hamiltonian~\eqref{RealCS} will exhibit two bands with different dispersion relations. One is an exact flat-band of $E_{k}=-t$, while the other is a dispersive band of $E_{k}=4J\cos(k)+t$~\cite{CDanieli2015}.

By applying dual transform $a_{n}=\frac{1}{\sqrt{N}}\sum_{k}a_{k}e^{-i2\pi\alpha kn}$ and $b_{n}=\frac{1}{\sqrt{N}}\sum_{k}b_{k}e^{-i2\pi\alpha kn}$ for Hamiltonian~\eqref{RealCS}, one can get the Hamiltonian in dual space, which has a similar structure to a Fano defect quasiperiodic lattice~[see Fig.~\ref{figcross}(b)], i.e.,
\begin{equation}\label{Momentumspace}
H_{K}=\sum_{k=1}^{N-1}(\lambda a_{k}^{\dagger}a_{k+1}+\mathrm{H.c.})+\sum_{k=1}^{N}[(t+J_{k}) a_{k}^{\dagger}b_{k}+\mathrm{H.c.}]+\sum_{k=1}^{N}J_{k}(a_{k}^{\dagger}a_{k}+b_{k}^{\dagger}b_{k}),
\end{equation}
where $J_{k}=2J \cos(2\pi\alpha k+\theta)$.

\subsubsection{A. The LE of lattice space}

The eigenequation set for Hamiltonian~\eqref{RealCS} is
\begin{equation}
\begin{aligned}\label{EqReal}
&\psi_{a,n-1}+\psi_{a,n+1}+\psi_{b,n-1}+\psi_{b,n+1}+t\psi_{b,n}=(E-V_{n})\psi_{a,n},\\
&\psi_{a,n-1}+\psi_{a,n+1}+\psi_{b,n-1}+\psi_{b,n+1}+t\psi_{a,n}=E\psi_{b,n}.
\end{aligned}
\end{equation}
One can get the relationship between $\psi_{a,n}$ and $\psi_{b,n}$ as $\psi_{b,n}=(E+t-V_{n})/(E+t)\psi_{a,n}$. Thus, from Eq.~\eqref{EqReal}, we obtain a new equation for the component $\psi_{a,n}$, i.e.,
\begin{equation}
\psi_{a,n+1}=\frac{E^2-2Et-t^2-EV_{n}}{2(E+t)-V_{n+1}}\psi_{a,n}-\frac{2(E+t)-V_{n-1}}{2(E+t)-V_{n+1}}\psi_{a,n-1},
\end{equation}
from which we directly obtain the corresponding transfer matrix
\begin{equation}\label{aer}
T_{n}=A_{n}B_{n},
\end{equation}
where
\begin{equation}
A_{n}=\frac{1}{2(E+t)-V_{n+1}},~B_{n}=\begin{pmatrix}
E^2-2Et-t^2-EV_{n} & -2(E+t)+V_{n-1} \\
2(E+t)-V_{n+1} &0
\end{pmatrix}.
\end{equation}

The LE can be written as $\gamma_{R}=\gamma_{A}+\gamma_{B}$, in which
\begin{equation}
\begin{aligned}
\gamma_{A}&=\lim_{N\rightarrow\infty}\frac{1}{N}\ln\prod_{n=1}^{N}\frac{1}{\left|2(E+t)-2\lambda\cos[2\pi\alpha (n+1)+\theta]\right|}\\
&=\frac{1}{2\pi}\int_{0}^{2\pi}\ln{\frac{1}{\left|2(E+t)+2\lambda\cos(\phi)\right|}}d\phi \\
&=\left\{\begin{matrix}
\ln\left|\frac{1}{|E+t|+\sqrt{(E+t)^2-\lambda^2}}\right|, &|E+t|>\lambda, \\
\ln|\frac{1}{\lambda }|, &|E+t|\le\lambda.
\end{matrix}\right.
\end{aligned}
\end{equation}
As for $\gamma_{B}$, we apply Avila's global theory of one-frequency analytical $SL(2,\mathbb{R} )$ cocycle~\cite{AAvila2015}. The first step is to perform an analytical continuation of the global phase $\theta\rightarrow\theta+i\epsilon$ in $B_{n}$. In large $\epsilon$ limit, one can get
\begin{equation}
B_{n,\epsilon\rightarrow\infty}=e^{-i2\pi\alpha n+\theta}e^{\epsilon}\begin{pmatrix}
-\lambda E & \lambda e^{i2\pi\alpha} \\
-\lambda e^{-i2\pi\alpha} &0
\end{pmatrix}+\mathcal{O}(1).
\end{equation}
According to Avila's global theory, as a function of $\epsilon$, $\gamma_{B}(E)$ is a convex piecewise linear function with integer slopes~\cite{AAvila2015}. The discontinuity of the slope occurs when $E$ belongs to the spectrum of Hamiltonian $H$ except for $\gamma_{B}(E) = 0$, which represents the extended states. One can get
\begin{equation}
\gamma_B(E)=\left\{\begin{matrix}
\ln\left|\frac{|\lambda E|+\sqrt{(\lambda E)^2-4\lambda^2}}{2}\right|, & |E|>2, \\
\ln|\lambda |, & |E|\le2.
\end{matrix}\right.
\end{equation}
Combining the information of $\gamma_{A}$ and $\gamma_{B}$, we obtain the LE $\gamma_R$ versus $E$ as
\begin{equation}
\begin{aligned}
\gamma_{R}=\left\{\begin{matrix}
\max\left\{\ln\left|\dfrac{|\lambda E|+\sqrt{(\lambda E)^2-4\lambda^2}}{|2(E+t)|+2\sqrt{(E+t)^2-\lambda^2}}\right|,0\right\}, &|E+t|>\lambda~\&~|E|>2, \\
\max\left\{\ln\left|\dfrac{|E|+\sqrt{E^2-4}}{2}\right|,0\right\}, &|E+t|\le\lambda~\&~|E|>2, \\
\max\left\{\ln\left|\dfrac{\lambda}{|E+t|+\sqrt{(E+t)^2-\lambda^2}}\right|,0\right\}, &|E+t|>\lambda~\&~|E|\le2,\\
0, &|E+t|\le\lambda ~\&~|E|\le2.
\end{matrix}\right.
\end{aligned}
\end{equation}
Since the second (third) row of the LE expression satisfies $|E|>2$ ($|E+t|<2$), then $\gamma_R>0$ ($\gamma_R=0$). Finally, one can obtain the LE in the lattice space as
\begin{equation}\label{crsgr}
\begin{aligned}
\gamma_{R}=\left\{\begin{matrix}
\max\left\{\ln\left|\dfrac{|\lambda E|+\sqrt{(\lambda E)^2-4\lambda^2}}{|2(E+t)|+2\sqrt{(E+t)^2-\lambda^2}}\right|,0\right\}, &|E+t|>\lambda~\&~|E|>2, \\
\ln\left|\dfrac{|E|+\sqrt{E^2-4}}{2}\right|, &|E+t|\le\lambda~\&~|E|>2, \\
0, &|E+t|>\lambda~\&~|E|\le2,\\
0, &|E+t|\le\lambda ~\&~|E|\le2.
\end{matrix}\right.
\end{aligned}
\end{equation}

\subsubsection{B. The LE of dual space}
The eigenequation set of dual Hamiltonian~\eqref{Momentumspace} is
\begin{equation}
\begin{aligned}
&\lambda(\psi_{a,k-1}+\psi_{a,k+1})+J_{k}\psi_{a,k}+(t+J_{k})\psi_{b,k}=E\psi_{a,k}, \\
&(t+J_{k})\psi_{a,k}+J_{k}\psi_{b,k}=E\psi_{b,k}.
\end{aligned}
\end{equation}
Then, the corresponding transfer matrix in dual space is
\begin{equation}
T_{k}=A_{k}B_{k},
\end{equation}
where
\begin{equation}
\begin{aligned}
&A_{k}=\dfrac{1}{E-2\cos(2\pi\alpha k+\theta)},\\
&B_{k}=\begin{pmatrix}
\dfrac{E^2-2(2E+2t)\cos(2\pi\alpha k+\theta)-t^2}{\lambda} & -E+2\cos(2\pi\alpha k+\theta) \\
E-2\cos(2\pi\alpha k+\theta) &0
\end{pmatrix}.
\end{aligned}
\end{equation}
The LE can be written as $\gamma_{R}(E)=\gamma_{A}(E)+\gamma_{B}(E)$, in which
\begin{equation}
\begin{aligned}
\gamma_{A}&=\lim_{N\rightarrow\infty}\frac{1}{N}\ln\prod_{k=1}^{N}\frac{1}{E-2\cos(2\pi\alpha k+\theta)}=\frac{1}{2\pi}\int_{0}^{2\pi}\ln{\frac{1}{\left| E+2\cos(\phi)\right|}}d\phi \\
&=\left\{\begin{matrix}
\ln\left|\frac{1}{|E|+\sqrt{(E)^2-4}}\right|, &|E+t|>\lambda, \\
0, &|E+t|\le\lambda.
\end{matrix}\right.
\end{aligned}
\end{equation}
As for $\gamma_{B}$, we apply Avila's global theory of one-frequency analytical $SL(2,\mathbb{R} )$ cocycle~\cite{AAvila2015}. The first step is to perform an analytical continuation of the global phase $\theta\rightarrow\theta+i\epsilon$ in $B_{k}$. In large $\epsilon$ limit, one can get
\begin{equation}
B_{k,\epsilon\rightarrow\infty}=e^{-i2\pi\alpha k+\theta}e^{\epsilon}\begin{pmatrix}
-(2E+2t) & 1 \\
-1 &0
\end{pmatrix}+\mathcal{O}(1).
\end{equation}
According to Avila's global theory, as a function of $\epsilon$, $\gamma_{B}(E)$ is a convex piecewise linear function with integer slopes~\cite{AAvila2015}. The discontinuity of the slope occurs when $E$ belongs to the spectrum of Hamiltonian $H$ except for $\gamma_{B}(E) = 0$, which represents the extended states. One can obtain
\begin{equation}
\gamma_B=\left\{\begin{matrix}
\ln\left|\dfrac{|E+t|+\sqrt{(E+t)^2-\lambda^2}}{\lambda}\right|, & |E+t|>\lambda, \\
0, & |E+t|\le\lambda.
\end{matrix}\right.
\end{equation}
Combining with $\gamma_{A}(E)$, the LE for different $E$ is
\begin{equation}
\gamma_{K}=\left\{\begin{matrix}
\max\left\{\ln\left|\dfrac{2(E+t)+2\sqrt{(E+t)^2-\lambda^2}}{\lambda(E+\sqrt{E^2-4})}\right|,0\right\}, &|E+t|>\lambda~\&~|E|>2, \\
\max\left\{\ln\left|\dfrac{2}{|E|+\sqrt{E^2-4}}\right|,0\right\}, &|E+t|\le\lambda~\&~|E|>2, \\
\max\left\{\ln\left|\dfrac{|E+t|+\sqrt{(E+t)^2-\lambda^2}}{\lambda}\right|,0\right\}, &|E+t|>\lambda~\&~|E|\le2,\\
0, &|E+t|\le\lambda~\&~|E|\le2.
\end{matrix}\right.
\end{equation}
Since the second (third) row of the LE satisfies $|E|>2$ ($|E+t|<2$), $\gamma_K=0$ ($\gamma_K>0$). Finally, one can obtain the LE of the lattice space as
\begin{equation}\label{crsgk}
\gamma_{K}=\left\{\begin{matrix}
\max\left\{\ln\left|\dfrac{2(E+t)+2\sqrt{(E+t)^2-\lambda^2}}{\lambda(E+\sqrt{E^2-4})}\right|,0\right\}, &|E+t|>\lambda~\&~|E|>2, \\
0, &|E+t|\le\lambda~\&~|E|>2, \\
\ln\left|\dfrac{|E+t|+\sqrt{(E+t)^2-\lambda^2}}{\lambda}\right|, &|E+t|>\lambda~\&~|E|\le2,\\
0, &|E+t|\le\lambda~\&~|E|\le2.
\end{matrix}\right.
\end{equation}

\subsubsection{C. Mobility edge}

By comparing the expressions of $\gamma_{R}$ and $\gamma_{K}$, one can find that in the region of $|E+t|\le\lambda~\&~|E|\le2$, $\gamma_R=\gamma_K=0$, indicating that the corresponding eigenstates in this region are multifractal critical states. For the rest of the energy region, the lattice and dual spaces have opposite localization properties, i.e., $\gamma_{R}=0$ while $\gamma_{K}>0$, or vice versa. Since the critical points satisfy a mirror-symmetric relationship between the case of $t<0$ and the case of $t>0$, here we will only exhibit the case of $t>0$ as what we have done in the main text.

\begin{figure}[thbp]
\centering
\includegraphics[width=13cm]{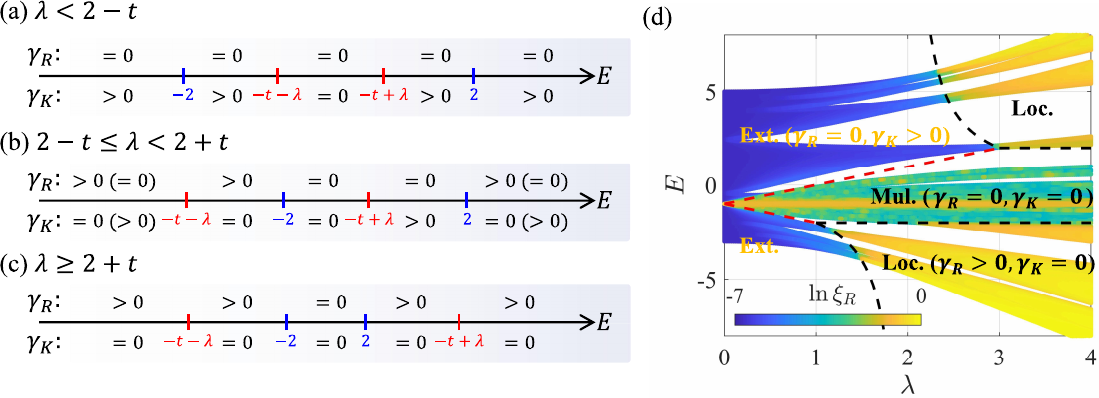}
\caption{(a-c) Phase diagram versus $E$ with different $\lambda$ for the case of $t<2$. (d) The lattice space IPR $\xi_{R}$ versus $\lambda$, where the black (red) dashed line is the critical energy separating $\gamma_{R}>0$ and $\gamma_{R}=0$ regions ($\gamma_{K}>0$ and $\gamma_{K}=0$ regions) in lattice (dual) space. The other parameters $N=610$ and $t=1$.}\label{CC1}
\end{figure}

On the one hand, for the case of $t<2$, from the inequalities in the analytic expressions~\eqref{crsgr} and~\eqref{crsgk}, one can find that the cross-stitch model has the same four critical points as the diamond model in the main text, and they also divide the energy axis into five intervals. The difference lies in that $E_{c}=\frac{2t}{\lambda-2}$ does not fall within the energy interval $|E+t|>\lambda$ and $|E|>2$ under the conditions of $\lambda<2-t$ and $\lambda\ge2+t$. That is to say, the three-state coexisting quantum phase can only emerge when $2-t\le\lambda<2+t$. In other words, since there is no localized (extended) state in the region of $\lambda< 2-t$ ($\lambda\ge 2+t$), only two-state coexisting quantum phases can emerge in certain circumstances [see Fig.~\ref{CC1}(a-c)]. Furthermore, we numerically compute the corresponding IPR [see Fig.~\ref{CC1}(d)]. The numerical results and analytical results show that MMEs and multi-state coexisting quantum phases will emerge in the system.

\begin{figure}[thbp]
\centering
\includegraphics[width=13cm]{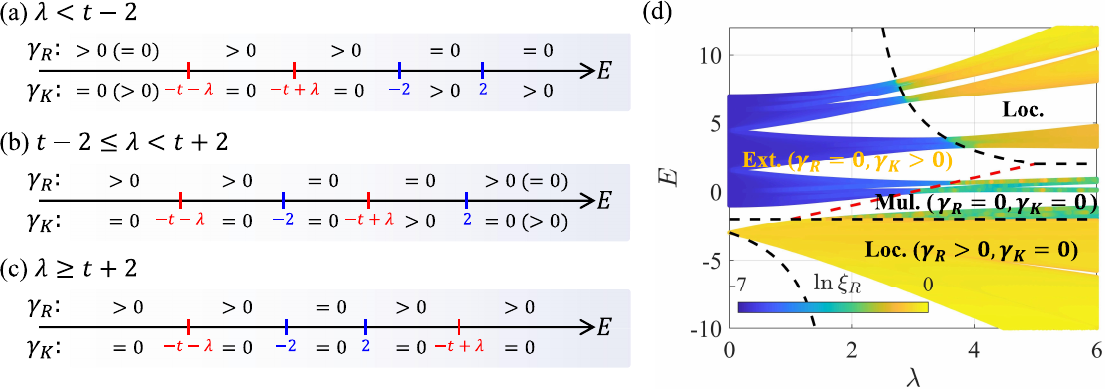}
\caption{(a-c) Phase diagram versus $E$ with different $\lambda$ for the case of $t\ge2$. (d) The lattice space IPR $\xi_{R}$ versus $\lambda$, where the black (red) dashed line is the critical energy separating $\gamma_{R}>0$ and $\gamma_{R}=0$ regions ($\gamma_{K}>0$ and $\gamma_{K}=0$ regions) in lattice (dual) space. The other parameters $N=610$ and $t=3$.}\label{CC3}
\end{figure}

A similar analysis leads us to the phase diagram for $t\ge2$, which is plotted in Fig.~\ref{CC3}(a-c). The main difference between $t\le2$ and $t<2$ is in the first stage of the phase diagram, i.e., the case of $\lambda<t-2$. Under certain circumstances, the four critical points satisfy the relative position relation $-t-\lambda<-t+\lambda<-2<2$. By analyzing the interval given by the inequalities of the expressions~\eqref{crsgr} and~\eqref{crsgk}, we find that it is impossible to have a region with multifractal states in this case. In other words, the system has only extended and localized states in this case. In the second stage ($2-t<\lambda<2+t$), multi-state coexisting quantum phases emerge in the system and all types of MMEs are allowed [see Fig.~\ref{CC3}(b)]. In the third stage ($\lambda\ge 2+t$), the system has only MMEs separating the localized state and the multifractal state [see Fig.~\ref{CC3}(c)]. Furthermore, we provide the numerical IPR [see Fig.~\ref{CC3}(d)], which is consistent with the conclusion given by the analytic expressions.

\begin{table*}[tbhp]
\renewcommand{\arraystretch}{2}
\centering
\caption{MMEs and quantum phases of partially-quasiperiodic cross-stitch lattice for the case of $t<2$}
\begin{tabular}{|c |c| c |c| c| c| c|}
\hline
Quasiperiodic strength & {$\lambda<2-t$} & \multicolumn{3}{c|}{$2-t\le\lambda<2+t$} & $\lambda\ge2+t$ \\ \hline
Exact MMEs ($E_c=$) & $\pm\lambda-t$ & $\frac{2t}{\lambda-2}$ & $\lambda-t$ & $-2$ & $\pm 2$ \\ \hline
Separated states & ~~Ext.$^*$ and Mul.~~ & Ext. and Loc. & Ext. and Mul. & Loc. and Mul. & Loc. and Mul. \\ \hline
Possible phases & {Ext.+Mul.} & \multicolumn{3}{c|}{Ext.+Mul.+Loc.} & Loc.+Mul. \\ \hline
\end{tabular}
\label{tabcorss}
\vspace{0.5em}

$^*$Ext.=Extended states; \quad Loc.=Localized states; \quad Mul.=Multifractal states.
\end{table*}
\begin{table*}[tbhp]
\renewcommand{\arraystretch}{2}
\centering
\caption{MMEs and quantum phases of partially-quasiperiodic cross-stitch lattice for the case of $t\ge2$}
\begin{tabular}{|c |c| c |c| c| c| c|}
\hline
Quasiperiodic strength & \multicolumn{2}{c|}{$\lambda< t-2$} & \multicolumn{3}{c|}{$2-t\le\lambda<2+t$} & $\lambda\ge2+t$ \\ \hline
Exact MMEs ($E_c=$) & $\frac{2t}{\lambda-2}$ & -2 & $\frac{2t}{\lambda-2}$ & $\lambda-t$ & -2 & $\pm 2$ \\ \hline
Separated states & Ext.$^*$ and Loc. & Ext. and Loc. & Ext. and Loc. & Ext. and Mul. & Ext. and Mul. & Loc. and Mul. \\ \hline
Possible phases & \multicolumn{2}{c|}{Ext.+Loc.} & \multicolumn{3}{c|}{Ext.+Mul.+Loc.} & Loc.+Mul. \\ \hline
\end{tabular}
\label{tabcorsst3}
\vspace{0.5em}
$^*$Ext.=Extended states; \quad Loc.=Localized states; \quad Mul.=Multifractal states.
\end{table*}

Finally, we summarize the MMEs and emergent quantum phases of the partially-quasiperiodic cross-stitch lattice in Tab.~\ref{tabcorss} and Tab.~\ref{tabcorsst3}. The results reveal that the system contains not only the traditional ME separating extended and localized states, but also the MMEs separating multifractal and localized states or separating multifractal and extended states. Meanwhile, exotic quantum phases featuring two-state coexistence (Ext.~+~Loc. or Loc.~+~Mul.) and three-state coexistence (Ext.~+~Mul.~+~Loc.) emerge.

This is another evidence supporting the conclusion: The MMEs and exotic quantum phases can emerge in the flat-band system.

\subsection{V-2. Quasiperiodic Lieb lattice}

\begin{figure}[thbp]
\centering
\includegraphics[width=15cm]{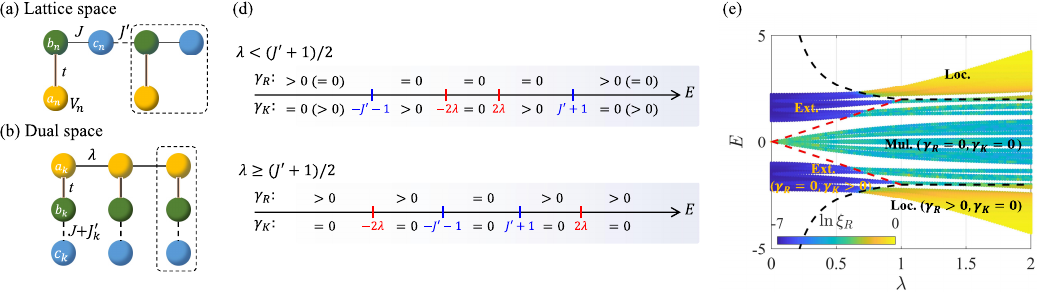}
\caption{(a) Lattice structure in real space and dual space. (b) The phase diagram of Lieb model. (C) IPR in lattice space $\xi_{R}$ versus $E$ for different $\lambda$, where the black (red) dashed line is the critical energy separating $\gamma_{R}(E) > 0$ ($\gamma_{K}(E) > 0$) and $\gamma_{R}(E) = 0$ ($\gamma_{K}(E) = 0$) in lattice (dual) space. In the computation, we set unit cell number $N=377$.}\label{Lieb}
\end{figure}

Now, we will show another quasiperiodic flat-band model, i.e., the quasiperiodic Lieb lattice~\cite{EHLieb1989}, and the schematic diagram is shown in Fig.~\ref{Lieb}(a). The corresponding Hamiltonian reads
\begin{equation}\label{eqstub}
H_{R}=\sum_{n=1}^{N}(ta_{n}^{\dagger}b_{n}+Jb_{n}^{\dagger}c_{n}+\mathrm{H.c.})+\sum_{n=1}^{N}(J'b_{n+1}^{\dagger}c_{n}+\mathrm{H.c.})+\sum_{n=1}^{N}V_{n}a_{n}^{\dagger}a_{n},
\end{equation}
where $V_{n}=2\lambda\cos(2\pi\alpha n+\theta)$. By means of the dual transform, one can obtain the Hamiltonian in dual space as
\begin{equation}\label{dualstub}
H_{K}=\sum_{k=1}^{N-1}\lambda( a_{k}^{\dagger}a_{k+1}+\mathrm{H.c.})+\sum_{k=1}^{N}[ta_{k}^{\dagger}b_{k}+(J+J'_{k})b_{k}^{\dagger}c_{k}+\mathrm{H.c.}],
\end{equation}
and the schematic diagram is shown in Fig.~\ref{Lieb}(b), where $J'_{k}=J' e^{i(2\pi\alpha k+\theta)}$. When $\lambda=0$, the Hamiltonian~\eqref{eqstub} has a flat band $E=0$. For the following discuss, we set $J=J'=t=1$ as the unit energy.

Note that, unlike the two flat-band models discussed earlier, the relative positions of the four critical points in this model are symmetric about the origin, so there are only two possibilities, namely $[-2,2]\subseteq[-J'-1,J'+1]$, or $[-J'-1,J'+1]\subseteq[-2,2]$. The obtained phase diagram is shown in Fig.~\ref{Lieb}(b). The corresponding IPR has also been given in Fig.~\ref{Lieb}(c).

\subsubsection{A. The LE of lattice space}
The eigenequation set for Hamiltonian~\eqref{eqstub} is
\begin{equation}
\begin{aligned}\label{realstub}
&V_{n}\psi_{a,n}+t\psi_{b,n}=E\psi_{a,n},\\
&J'\psi_{c,n-1}+t\psi_{a,n}+\psi_{c,n}=E\psi_{b,n},\\
&\psi_{b,n}+J'\psi_{b,n+1}=E\psi_{c,n}.
\end{aligned}
\end{equation}
Then, by simplifing the Eq.~\eqref{realstub}, one can get
\begin{equation}
\psi_{c,n+1}=\frac{(E^2-J'^2-1)(E-V_{n})-Et^2}{J'(E-V_{n})}\psi_{c,n}-\psi_{c,n-1}.
\end{equation}
Then, one can directly obtain the corresponding transfer matrix
\begin{equation}
T_{n}=A_{n}B_{n},
\end{equation}
where
\begin{equation}
\begin{aligned}
&A_{n}=\frac{1}{J'[E-2\lambda\cos(2\pi\alpha n+\theta)]},\\
&B_{n}=\begin{pmatrix}
(E^2-J'^2-1)(E-V_{n})-Et^2 &-J'(E-V_{n}) \\
J'(E-V_{n}) &0
\end{pmatrix}.
\end{aligned}
\end{equation}
The LE can be written as $\gamma_{R}(E)=\gamma_{A}(E)+\gamma_{B}(E)$, in which
\begin{equation}
\begin{aligned}
\gamma_{A}(E)&=\lim_{N\rightarrow\infty}\frac{1}{N}\ln\prod_{n=1}^{N}\frac{1}{J'[E-2\lambda\cos(2\pi\alpha n+\theta)]}\\
&=\left\{\begin{matrix}
\ln\left|\dfrac{2}{J'(|E|+\sqrt{E^2-4\lambda^2})}\right|, &|E|>2\lambda, \\
\ln|\dfrac{1}{J'\lambda}|, &|E|\le 2\lambda.
\end{matrix}\right.
\end{aligned}
\end{equation}
For $\gamma_{B}(E)$, again, we apply Avila's global theory of one-frequency analytical $SL(2,\mathbb{R} )$ cocycle~\cite{AAvila2015}. The first step in the calculation is to perform an analytical continuation of the global phase $\theta\rightarrow\theta+i\epsilon$ in $B_{n}$. In large $\epsilon$ limit, one can obtain
\begin{equation}
B_{n,\epsilon\rightarrow\infty}=e^{-i2\pi\alpha n+\theta}e^{\epsilon}\begin{pmatrix}
-\lambda(E^2-J'^2-1) & J'\lambda \\
-J'\lambda &0
\end{pmatrix}+\mathcal{O}(1).
\end{equation}
According to Avila's global theory, as a function of $\epsilon$, $\gamma_{B}(E)$ is a convex piecewise linear function with integer slopes~\cite{AAvila2015}. The discontinuity of the slope occurs when $E$ belongs to the spectrum of Hamiltonian $H$ except for $\gamma_{B}(E) = 0$, which represents the extended states. Then, one can get
\begin{equation}
\gamma_B=\left\{\begin{matrix}
\ln\left|\lambda\frac{|E^2-J'^2-1|+\sqrt{(E^2-J'^2-1)^2-4J'^2}}{2}\right|, & |E^2-J'^2-1|>2J', \\
\ln|J'\lambda|, & |E^2-J'^2-1|\le 2J'.
\end{matrix}\right.
\end{equation}
Combining with $\gamma_{A}$, the LE for different $E$ is
\begin{equation}
\gamma_{R}=\left\{\begin{matrix}
\max\left\{\ln\left|\dfrac{\lambda|E^2-J'^2-1|+\lambda\sqrt{(E^2-J'^2-1)^2-4J'^2}}{J'(|E|+\sqrt{E^2-4\lambda^2})}\right|,0\right\}, &|E|>2\lambda~\&~|E^2-J'^2-1|>2J', \\
\max\left\{\ln\left|\dfrac{|E^2-J'^2-1|+\sqrt{(E^2-J'^2-1)^2-4J'^2}}{2J'}\right|,0\right\}>0, &|E|\le 2\lambda~\&~|E^2-J'^2-1|>2J', \\
\max\left\{\ln\left|\dfrac{2\lambda}{|E|+\sqrt{E^2-4\lambda^2}}\right|,0\right\}=0, &|E|>2\lambda~\&~|E^2-J'^2-1|\le 2J',\\
0, &|E|\le 2\lambda~\&~|E^2-J'^2-1|\le 2J'.
\end{matrix}\right.
\end{equation}
For the energy interval in the second (third) row, since $|E^2-J'^2-1|>2J'$ ($|E|>2\lambda$), it follows that $\gamma_R>0$ ($\gamma_R=0$). Therefore, the final LE of the system in real space is
\begin{equation}\label{realgLieb}
\gamma_{R}=\left\{\begin{matrix}
\max\left\{\ln\left|\dfrac{\lambda|E^2-J'^2-1|+\lambda\sqrt{(E^2-J'^2-1)^2-4J'^2}}{J'(|E|+\sqrt{E^2-4\lambda^2})}\right|,0\right\}, &|E|>2\lambda~\&~|E^2-J'^2-1|>2J', \\
\ln\left|\dfrac{|E^2-J'^2-1|+\sqrt{(E^2-J'^2-1)^2-4J'^2}}{2J'}\right|, &|E|\le 2\lambda~\&~|E^2-J'^2-1|>2J', \\
0, &|E|>2\lambda~\&~|E^2-J'^2-1|\le 2J',\\
0, &|E|\le 2\lambda~\&~|E^2-J'^2-1|\le 2J'.
\end{matrix}\right.
\end{equation}

\subsubsection{B. The LE of dual space}
The eigenequation set for dual space Hamiltonian~\eqref{dualstub} is
\begin{equation}
\begin{aligned}
& t\psi_{b,k}+\lambda\psi_{a,k-1}+\lambda\psi_{a,k+1}=E\psi_{a,k},\\
& t\psi_{a,k}+(J+J' e^{i(2\pi\alpha k+\theta)})\psi_{c,k}=E\psi_{b,k},\\
& (J+J' e^{-i(2\pi\alpha k+\theta)})\psi_{b,k}=E\psi_{c,k}.
\end{aligned}
\end{equation}
Similarly, by simplifing the eigenequation set, one can get
\begin{equation}
\psi_{a,n+1}=\frac{E^3-E(2J'-1-t^2)-2EJ'\cos(2\pi\alpha k+\theta)}{\lambda[E^2-J'^2-1-2J'\cos(2\pi\alpha k+\theta)]}\psi_{a,n}-\psi_{a,n-1}
\end{equation}
Then, one can directly obtain the corresponding transfer matrix
\begin{equation}
\begin{aligned}
T_{k}=A_{k}B_{k},
\end{aligned}
\end{equation}
where
\begin{equation}
\begin{aligned}
&A_{k}=\frac{1}{\lambda[E^2-J'^2-1-2J'\cos(2\pi\alpha k+\theta)]},\\
&B_{k}=\begin{pmatrix}
E^3-E(2J'-1-t^2)-2EJ'\cos(2\pi\alpha k+\theta) & -\lambda[E^2-J'^2-1-2J'\cos(2\pi\alpha k+\theta)] \\
\lambda[E^2-J'^2-1-2J'\cos(2\pi\alpha k+\theta)] &0
\end{pmatrix}.
\end{aligned}
\end{equation}
Then, LE can be written as $\gamma_{R}(E)=\gamma_{A}(E)+\gamma_{B}(E)$, where
\begin{equation}
\begin{aligned}
\gamma_{A}&=\lim_{N\rightarrow\infty}\frac{1}{N}\ln\prod_{k=1}^{N}\frac{1}{\lambda[E^2-J'^2-1-2J'\cos(2\pi\alpha k+\theta)]}\\
&=\left\{\begin{matrix}
\ln\left|\dfrac{2}{\lambda(|E^2-J'^2-1|+\sqrt{(E^2-J'^2-1)^2-4J'^2})}\right|, &|E^2-J'^2-1|>2J', \\
\ln|\dfrac{1}{J'\lambda}|, &|E^2-J'^2-1|\le 2J'.
\end{matrix}\right.
\end{aligned}
\end{equation}
As for $\gamma_{B}$, by applying Avila's global theory, one can get
\begin{equation}
\gamma_B=\left\{\begin{matrix}
\ln\left|\dfrac{|EJ'|+J'\sqrt{(E)^2-4\lambda^2}}{2}\right|, & |E|>2\lambda, \\
\ln|J'\lambda|, & |E|\le 2\lambda.
\end{matrix}\right.
\end{equation}
Combining with $\gamma_{A}$, the LE for different $E$ is
\begin{equation}
\gamma_{K}=\left\{\begin{matrix}
\max\left\{\ln\left|\dfrac{J'(|E|+\sqrt{E^2-4\lambda^2})}{\lambda|E^2-J'^2-1|+\lambda\sqrt{(E^2-J'^2-1)^2-4J'^2}}\right|,0\right\}, &|E|>2\lambda~\&~|E^2-J'^2-1|>2J', \\
\max\left\{\ln\left|\dfrac{2J'}{|E^2-J'^2-1|+\sqrt{(E^2-J'^2-1)^2-4J'^2}}\right|,0\right\}, &|E|\le 2\lambda~\&~|E^2-J'^2-1|>2J', \\
\max\left\{\ln\left|\dfrac{|E|+\sqrt{E^2-4\lambda^2}}{2\lambda}\right|,0\right\}, &|E|>2\lambda~\&~|E^2-J'^2-1|\le 2J',\\
0, &|E|\le 2\lambda~\&~|E^2-J'^2-1|\le 2J'.
\end{matrix}\right.
\end{equation}
Similarly, since the second (third) line of inequality satisfies $|E^2-J'^2-1|>2J'$ ($|E|>2\lambda$), we have $\gamma_K=0$ ($\gamma_K>0$). The final LE for dual space is
\begin{equation}\label{momengLieb}
\gamma_{K}=\left\{\begin{matrix}
\max\left\{\ln\left|\dfrac{J'(|E|+\sqrt{E^2-4\lambda^2})}{\lambda|E^2-J'^2-1|+\lambda\sqrt{(E^2-J'^2-1)^2-4J'^2}}\right|,0\right\}, &|E|>2\lambda~\&~|E^2-J'^2-1|>2J', \\
0, &|E|\le 2\lambda~\&~|E^2-J'^2-1|>2J', \\
\ln\left|\dfrac{|E|+\sqrt{E^2-4\lambda^2}}{2\lambda}\right|, &|E|>2\lambda~\&~|E^2-J'^2-1|\le 2J',\\
0, &|E|\le 2\lambda~\&~|E^2-J'^2-1|\le 2J'.
\end{matrix}\right.
\end{equation}

\subsubsection{C. Mobility edge}
Comparing the LE in the two dual spaces, one can find that the eigenstates in the region of $|E|\le2\lambda~\&~|E^2-J'^2-1|\le2J'$ are delocalized in both spaces, which means they are actually the multifractal state. Moreover, the LE does not depend on the coupling parameter $t$. In other words, for arbitrarily small $t$, one can induce multifractal states in this lattice model. The ME between the extended and localized states is determined by the LE of the energy region with $|E|>2\lambda~\&~|E^2-J'^2-1|>2J'$ for $E_{c}=\frac{J'\pm\sqrt{J'^2+4\lambda^2(J'^2+1)}}{2\lambda}$.

\begin{table*}[tbhp]
\renewcommand{\arraystretch}{2}
\centering
\caption{MMEs and quantum phases of partially-quasiperiodic lieb lattice}
\begin{tabular}{|c |c| c | c|}
\hline
Quasiperiodic strength & \multicolumn{2}{c|}{$\lambda<(J'+1)/2$} & {$\lambda\ge(J'+1)/2$} \\ \hline
Exact MMEs ($E_c=$) & $\pm2\lambda$ & $\frac{J'\pm\sqrt{J'^2+4\lambda^2(J'^2+1)}}{2\lambda}$ & $\pm2$ \\ \hline
Separated states & Ext.$^*$ and Mul. & Ext. and Loc. & Ext. and Mul. \\ \hline
Possible phases & \multicolumn{2}{c|}{Ext.+Loc.,~Ext+Mul.+Loc.} & {Loc.+Mul.} \\ \hline
\end{tabular}
\label{tablieb}
\vspace{0.5em}

$^*$Ext.=Extended states; \quad Loc.=Localized states; \quad Mul.=Multifractal states.
\end{table*}

The LE versus $E$ and $\lambda$ is shown in Fig.~\ref{Lieb}(b).

In the first stage [$\lambda<(J'+1)/2$], the system has four MEs. Two are $E_c=\pm \lambda$ separating the extended and the multifractal state, while the other two are $E=\frac{J'\pm\sqrt{J'^2+4\lambda^2(J'^2+1)}}{2\lambda}$ separating the extended and the localized state.

In other words, though MMEs separating multifractal and localized states are not found in the first stage, this does not mean that the three-state coexisting quantum phase can not appear. As shown in Fig.~\ref{Lieb}(c), the IPR reflects that near $\lambda = 0.7$, three-state coexisting quantum phase emerges.

In the second stage [$\lambda\ge(J'+1)/2$], the region $-J'-1\le\frac{J'\pm\sqrt{J'^2+4\lambda^2(J'^2+1)}}{2\lambda}\le J'+1$ no longer satisfies the energy interval in the first line of Eq.~\eqref{realgLieb} and~\eqref{momengLieb}. As a result, the MEs separating the extended and localized states disappear. At this point the system enters a quantum phase in which the multifractal and the localized state coexist, separated by MMEs of $E=\pm2$.

Finally, we summarize the MMEs and emergent quantum phases of the partially-quasiperiodic Lieb lattice in Tab.~\ref{tablieb}. The results again reveal that the system contains not only the traditional ME separating extended and localized states, but also the MMEs separating multifractal and localized states or separating multifractal and extended states. Meanwhile, exotic quantum phases featuring two-state coexistence (Ext.~+~Loc. or Loc.~+~Mul.) and three-state coexistence (Ext.~+~Mul.~+~Loc.) emerge.

This is yet another evidence supporting the conclusion: The MMEs and exotic quantum phases can emerge in the flat-band system.

\section{VI. Experimental scheme of the Rydberg atomic array}

\begin{figure}[htbp]
	\centering	
    \includegraphics[width=16cm]{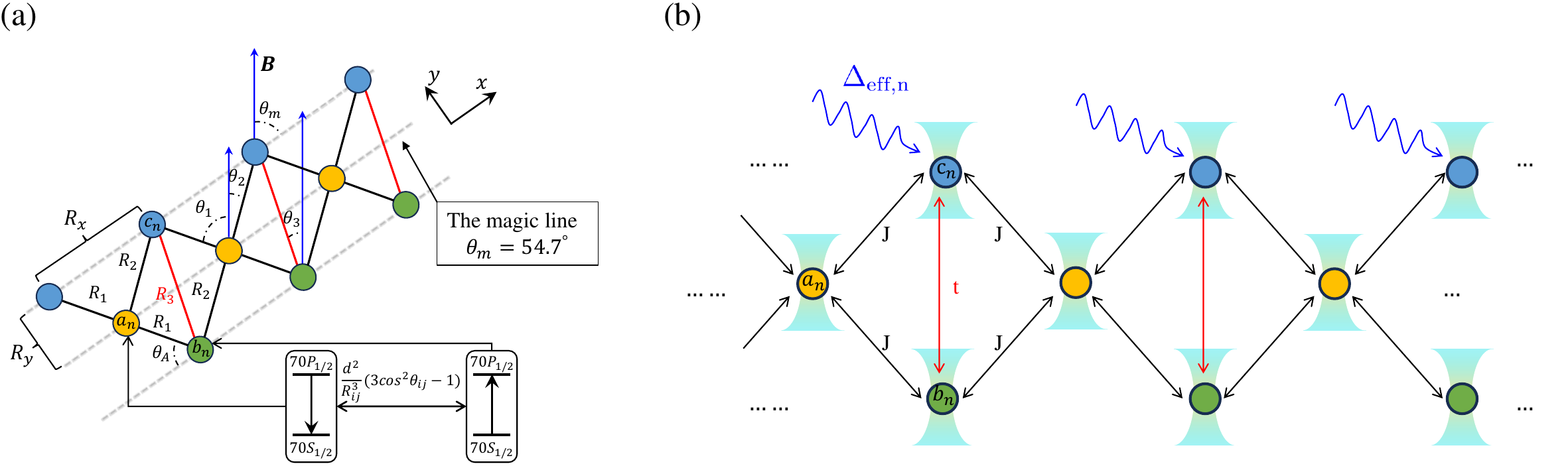}
	\caption{Experimental scheme on MMEs of model~\eqref{eqdim} in Rydberg atomic array. The corresponding angles are marked.}\label{ExpSetup}
\end{figure}

Experimentally, one can realize the MMEs in diamond flat-band model by the following spin Hamiltonian
\begin{equation}\label{spinH}
\begin{aligned}
    H_{s}=&\sum_{j_{x}}(J\sigma _{j_{x},A}^{+}\sigma_{j_{x},B}^{-}+J\sigma _{j_{x},A}^{+}\sigma_{j_{x},C}^{-}+t\sigma _{j_{x},B}^{+}\sigma_{j_{x},C}^{-}+J\sigma _{j_{x},B}^{+}\sigma_{j_{x}+1,A}^{-}+J\sigma _{j_{x},C}^{+}\sigma_{j_{x}+1,A}^{-}+\mathrm{H.c.} )\\
    &+\frac{1}{2}\sum_{j_{x}}V_{j_x}(\mathbb{I}+\sigma_{j_{x},C}^{z} ).
\end{aligned}
\end{equation}
This Hamiltonian can be transformed to the diamond quasiperiodic lattice by relabeling site index $j_{x}\rightarrow n$ for each leg and defining operator $b^{+}_{n}=\left|\uparrow \right>_{n}\left<\downarrow \right|_{n}$ at each site. We consider the three-legged superarray of Rydberg atoms, and each atom is trapped in optical tweezers. The schematic diagram is shown in Fig.~\ref{ExpSetup}(a), where $R_{1}=\frac{2Ry}{\sin{\theta_A}}$, $R_2=\sqrt{R_{1}^{2}+R_{x}^{2}-2R_{1}R_{x}\cos{\theta_A}}$, $R_3=\sqrt{2R_{1}R_{2}\cos{(\theta_{1}+\theta_{2})}}$. From the sine and cosine theorems, we can further give the angle between the sublattice $\theta_{1}=\pi-\theta_m-\theta_A$, $\theta_{2}=\theta_m-\arcsin{(\frac{R_1}{R_2}\sin{\theta_A})}$ and $\theta_{3}=\theta_m-\arcsin{(\frac{2R_{1}}{R_3}\sin{\theta_A})}$, where $\theta_m=54.7^{\circ}$ is the magic angle.

 To realize the above Hamiltonian~\eqref{spinH} in Rydberg atomic array, angle-dependent dipole-dipole interactions and the AC Stark potential need to be realized successively, then the total Hamiltonian of the experimental system reads
\begin{equation}
    H=H_{dipole}+H_{AC}(r),
\end{equation}
where the AC Stark term reads
\begin{equation}
    H_{AC}=\sum_{j_{x}}\frac{|\Omega_0|^2}{\Delta}\cos^2(2\pi\alpha j_{x})(\mathbb{I}+\sigma_{j_{x},C}^{z}),
\end{equation}
and the dipole-dipole term reads
\begin{equation}
\begin{aligned}
H_{dipole}=&\sum_{j_{x}}(J_{AB}\sigma _{j_{x},A}^{+}\sigma_{j_{x},B}^{-}+J_{AC}\sigma _{j_{x},A}^{+}\sigma_{j_{x},C}^{-}+J_{BC}\sigma _{j_{x},B}^{+}\sigma_{j_{x},C}^{-}+J_{BA}\sigma _{j_{x},B}^{+}\sigma_{j_{x}+1,A}^{-}+J_{CA}\sigma _{j_{x},C}^{+}\sigma_{j_{x}+1,A}^{-}+\mathrm{H.c.} )\\
&+\sum_{|i-j|>1}(\frac{J_{ij}}{R_{ij}^{3}}\sigma^{+}_{i,A}\sigma_{j,B}^{-}+\frac{J_{ij}}{R_{ij}^{3}}\sigma^{+}_{i,A}\sigma_{j,C}^{-}+\frac{J_{ij}}{R_{ij}^{3}}\sigma^{+}_{i,B}\sigma_{j,C}^{-}+\mathrm{H.c.}).
\end{aligned}
\end{equation}
Here $\sigma^{\pm} = \frac{1}{2}(\sigma_x \pm i\sigma_y)$, $\sigma_x$ and $\sigma_y$ are the  Pauli matrices. The dipole-dipole interaction between Rydberg atoms is given by $J_{ij} = \frac{d^2}{R_{ij}^3}(3\cos^2\theta_{ij} - 1),$ where $d$ represents the transition dipole moment between the two Rydberg levels, $R_{ij}$ with $i, j = A, B, C$ is the distance between sites \(i\) and $j$, and $\theta_{ij}$ is the angle between $R_{ij}$ and the quantization axis defined by the magnetic field \(\mathbf{B}\)~\cite{Sdeleseleuc2019}. \(J_{ii}\) can be effectively mitigated to zero by selecting the angle \(\theta_{ii}\) to be the magic angle $\theta_m = 54.7^{\circ}$. Note that, the hoping term $J_{AB}=J_{AC}=J_{BA}=J_{CA}=J$ and $J_{BC}=t$, i.e., $\frac{d^2}{R_{1}^{3}}(3\cos^{2}\theta_{1}-1)=\frac{d^2}{R_{2}^{3}}(3\cos^{2}\theta_{2}-1)=J$ and $\frac{d^2}{R_{3}^{3}}(3\cos^{2}\theta_{3}-1)=t$. The non-nearest neighboring term can  be safely ignored since its value decays with distance $R_{ij}^3$. Then, by setting values of any two of $\theta_A,~R_x$ and $R_y$, the value of the third can be readily obtained by expressions $\frac{d^2}{R_{1}^{3}}(3\cos^{2}\theta_{1}-1)=\frac{d^2}{R_{2}^{3}}(3\cos^{2}\theta_{2}-1)$. For example, if we give $\theta_A=50^{\circ}$, $R_x=0.5a$, we get $R_y=0.652a$. Then we obtain $J\propto 3.59d^2$ and $t\propto 0.66d^2$. In this case, we have $t/J=0.184$.



\section{VII. The measurement scheme}

\subsection{VII-1. Basic principles}

{In this section, we discuss the measurement scheme of MMEs. We here address an experimentally feasible method to detect the IPRs and LEs based on a powerful spectroscopic approach outlined in Ref.~\cite{PRoushan2017}. }

{According to the basic principles of quantum mechanics, the dynamics of a quantum system with time-independent Hamiltonian satisfies the Schr\"{o}dinger equation. One can get the wave function versus time, i.e.,
\begin{equation}\label{Evo}
|\psi(t)\rangle=e^{-iHt}|\psi(0)\rangle=\sum_{\beta}e^{-iHt}|\psi_{\beta}\rangle \left \langle \psi_{\beta}  | \psi(0)  \right \rangle =\sum_{\beta}C_{\beta}e^{-iE_{\beta}t}|\psi_\beta\rangle,
\end{equation}
where $C_{\beta}=\left\langle\psi_{\beta} | \psi(0)  \right \rangle $ and $\beta\in\{{1,2,3,\cdots,3N}\}$ corresponds to the eigenvalue index. One can obtain the expected value of an observable $\hat{O}$ at time $t$ as
\begin{equation}
O(t)=\langle\psi(t)| \hat{O}|\psi(t)\rangle=\sum_{\beta, \beta^{\prime}} O_{\beta^{\prime}, \beta} C_{\beta} C_{\beta^{\prime}}^{*} e^{-i\left(E_{\beta}-E_{\beta^{\prime}}\right) t},
\end{equation}
where $O_{\beta',\beta}=\langle\psi_{\beta'}|\hat{O}|\psi_{\beta}\rangle$. From the expression, one can find that to get the expected value of an observable, we need to measure the energy difference. Experimentally, we use the spectroscopic method proposed in Ref.~\cite{PRoushan2017} to extract the eigenvalues. In other words, what we want to measure is $E_{\beta}$ rather than $E_{\beta}-E_{\beta^{'}}$. This problem can be solved by fixing $E_{\beta^{'}}$ and using it as a reference energy~\cite{PRoushan2017}. Specifically, since the manifold that contains the vacuum state is a manifold containing only one state, we can extract the eigenenergy $E_{\beta}$ by choosing the superposition of the vacuum state and the single excited state as the initial state, i.e.,
\begin{equation}\label{expint}
|\psi(0)\rangle_m=|0\rangle_1 \cdots|0\rangle_{m-1} (\frac{|0\rangle_m+|1\rangle_m}{\sqrt{2}}) |0\rangle_{m+1} \cdots|0\rangle_{3N}=\frac{1}{\sqrt{2}}(\left |\mathrm{Vac}  \right \rangle +\left |\mathbf{1}_{m}   \right \rangle ),
\end{equation}
where $\left |\mathrm{Vac}  \right \rangle$ is the vacuum state and $\left |\mathbf{1}_{m} \right \rangle$ is an excited state at the $m$-th site. Under such circumstances, the corresponding Eq.~\eqref{Evo} can be rewritten as
\begin{equation}
    |\psi(t)\rangle_{m}=e^{-iHt}|\psi(0)\rangle_{m}=\frac{1}{\sqrt{2}}\left(\left |\mathrm{Vac}  \right \rangle +\sum_{\beta}C_{\beta,m}e^{-iE_{\beta}t}\left|\psi_{\beta}   \right \rangle  \right),
\end{equation}
where $C_{\beta,m}=\left \langle \psi_{\beta}  | \mathrm{1}_{m}  \right \rangle$ and $\left | \psi_{\beta}  \right \rangle$ is an eigenstate of a singly excited state with the eigenenergy $E_{\beta}$. To measure the eigenvalue $E_{\beta}$, we need to measure the evolution of $\langle\sigma^x_m\rangle$ and $\langle\sigma^y_m\rangle$ to construct the annihilation operator, i.e., $\left \langle \hat{a} \right \rangle =\langle\sigma^+_m\rangle$, where $\langle\sigma^+_m\rangle =\langle\sigma^x_m\rangle+i\langle\sigma^y_m\rangle$. Then, one can get the expected value of $\sigma_{m}^{+}$, i.e.,
\begin{equation}\label{Cp}
  \langle \sigma^+_m\rangle=\frac{1}{2}\sum_{\beta}|C_{\beta,m}|^2e^{-iE_{\beta}t}.
\end{equation}
The complete basis vectors are formed by scaning the parameter $m$ from 1 to $3N$, such that each eigenstate has an overlap with all the different single excited initial states. In turn, each single excited initial state can expand by the eigenstates' complete basis, the coefficient of which is the probability that the initial state projects on different eigenstates.}
\begin{figure*}[hp]
\centering
\includegraphics[width=10cm]{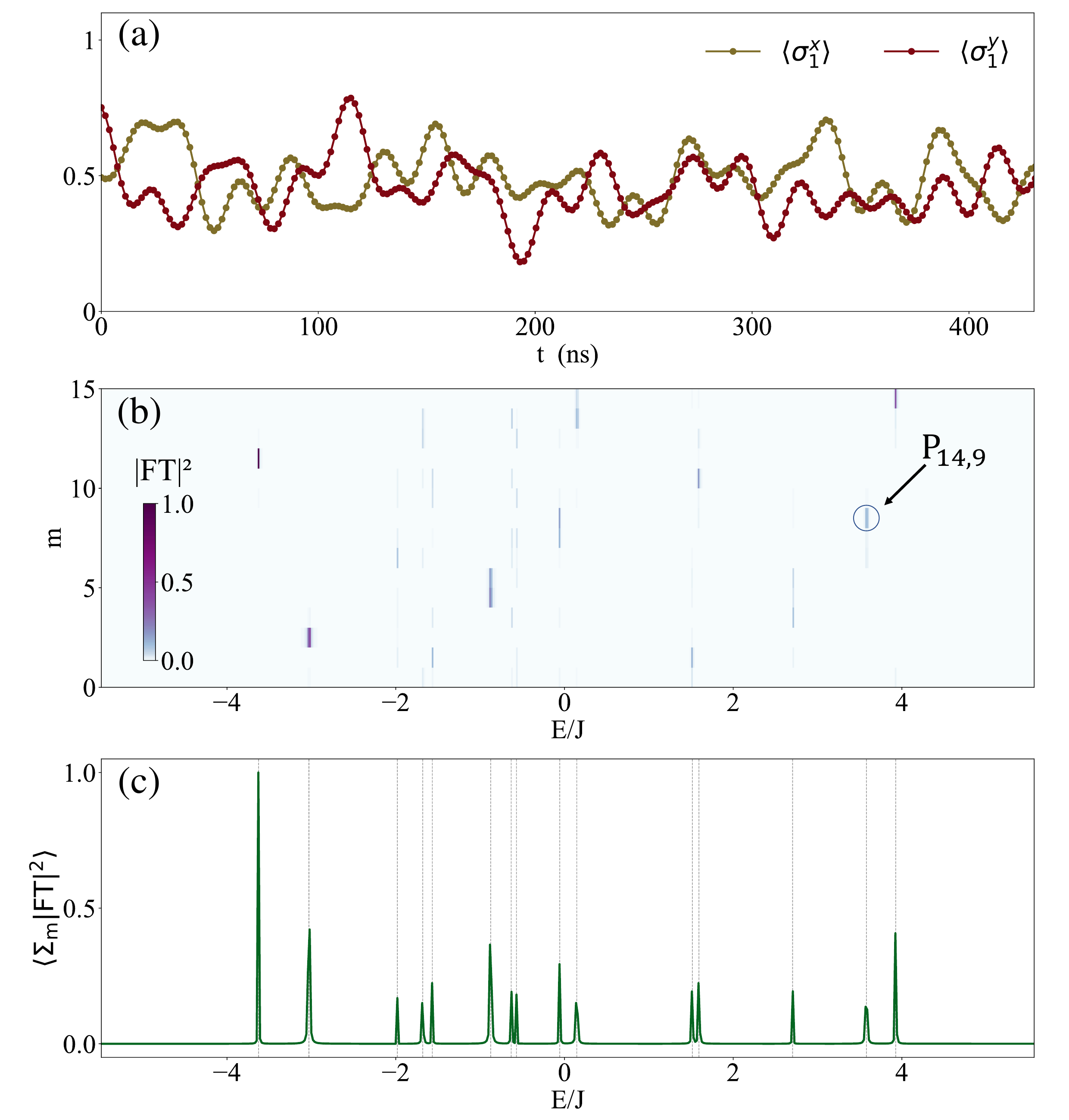}
\caption{(a) Typical data set showing $\langle\sigma^{x}_1\rangle$ and $\langle\sigma^{y}_1\rangle$ versus time. (b) The Fourier transformation (FT) of $\langle\sigma^+_m\rangle$ curves. $m\in\{1,2,\cdots,15\}$ correspond to different initial states. The probability of a single spin-flip state on the $9$-th site in the $14$-th eigenstate $\text{P}_{14,11}$ is highlighted. (c) Average |FT|$^2$ amplitudes of the data in (b). fifteen peaks emerge. Throughout, $J=2\pi\times7.53$MHz.}\label{ms2}
\end{figure*}

{Note that, Eq.~\eqref{Cp} is structurally identical to the expression of discrete Fourier transform. Then, based on the similarity of the two, the eigenenergy and eigenstate can be extracted by Fourier transform (FT) from the  evolution process. Specifically, the frequency in the Fourier transform expression corresponds to the eigenenergy $\{E_{\beta}\}$, and the amplitude of FT corresponds to $\{C_{\beta}\}$.}

 {
 Now, we use a $5$-unit-cell (15 Rydberg atoms) Rydberg atomic array as an example to briefly describe the measurement process.
 During the evolution process, one can obtain the time evolution curve of $\langle\sigma^x_m\rangle$ and $\langle\sigma^y_m\rangle$ [see Fig.~\ref{ms2}(a)]. By conducting a Fourier transform on the evolution curve of $\langle\sigma^+_m\rangle$, one can obtain eigenenergy $E_{\beta}$ and the corresponding modular square $|\text{FT}|^2$ on each site [see Fig.~\ref{ms2}(a)], which can be defined as $\text{P}_{\beta,m}^{R}$ [see Fig.~\ref{ms2}(b)]. In view of the correspondence relation between Eq.~\eqref{Cp} and the expression of Fourier transform, one can find that $\text{P}_{\beta,m}^{R}$ is actually the density distribution on $m$-th site. After obtaining all the $\text{P}_{\beta,m}^{R}$'s values, one can directly obtain the corresponding IPR $\xi_R$ through the following relation
\begin{equation}\label{expIPR}
    \xi_R(E_{\beta})=\sum_n(\text{P}_{\beta,m}^{R})^2.
\end{equation}
Furthermore, by allowing such single site spin-flip to traverse the entire atomic chain from $m=1$ to $m=15$, one can experimentally extract all eigenvalues of the system~[see Fig.~\ref{ms2}(c)].}

\begin{figure*}[thbp]
\centering
\includegraphics[width=8.5cm]{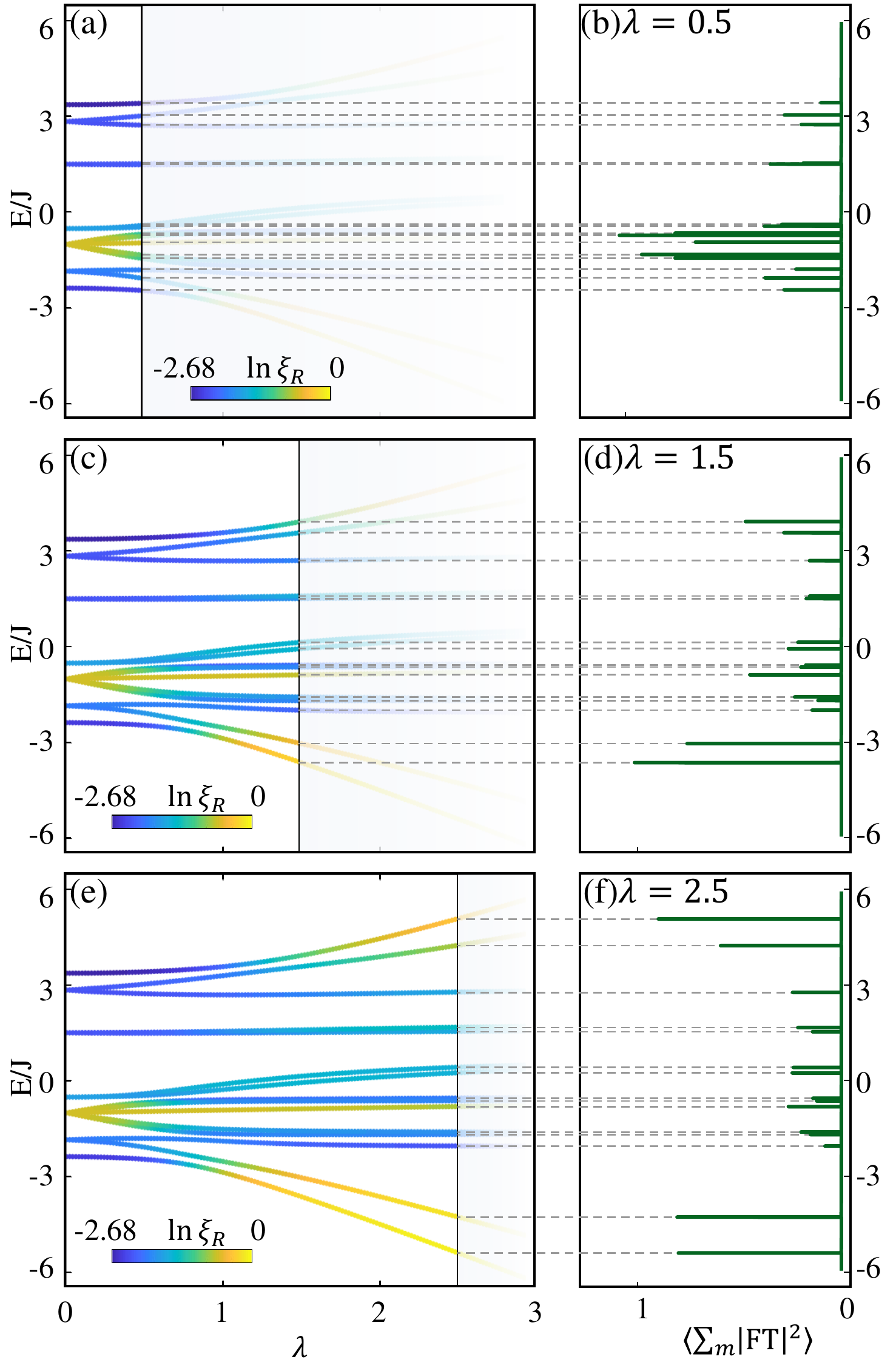}
\caption{The correspondence between the theoretically calculated eigenvalues (the left column) of different quasiperiodic strengths in the IPR diagram and the eigenvalues extracted from numerical simulated experiments(the right column). Throughout, $J=2\pi\times7.53$MHz.}\label{ms3}
\end{figure*}

{By changing the quasiperiodic strength $\lambda$, we can obtain the phase diagram in lattice space. We numerically simulate the process of extracting IPR and draw the phase diagram. The results are plotted in  Fig.~\ref{ms3} for the case of $5$ unit cells. We compare the eigenenergy predicted by the theory (left column in Fig.~\ref{ms3}) in the main text with the eigenvalues extracted by the spectroscopic method for quasiperiodic strength $\lambda=0.5,~1.5,~2.5$ (right column in Fig.~\ref{ms3}). The results agree well with each other.}

\subsection{VII-2. The comparasion of IPRs obtained by the diagonalization and spectroscopic method}

\begin{figure*}[thbp]
\centering
\includegraphics[width=17cm]{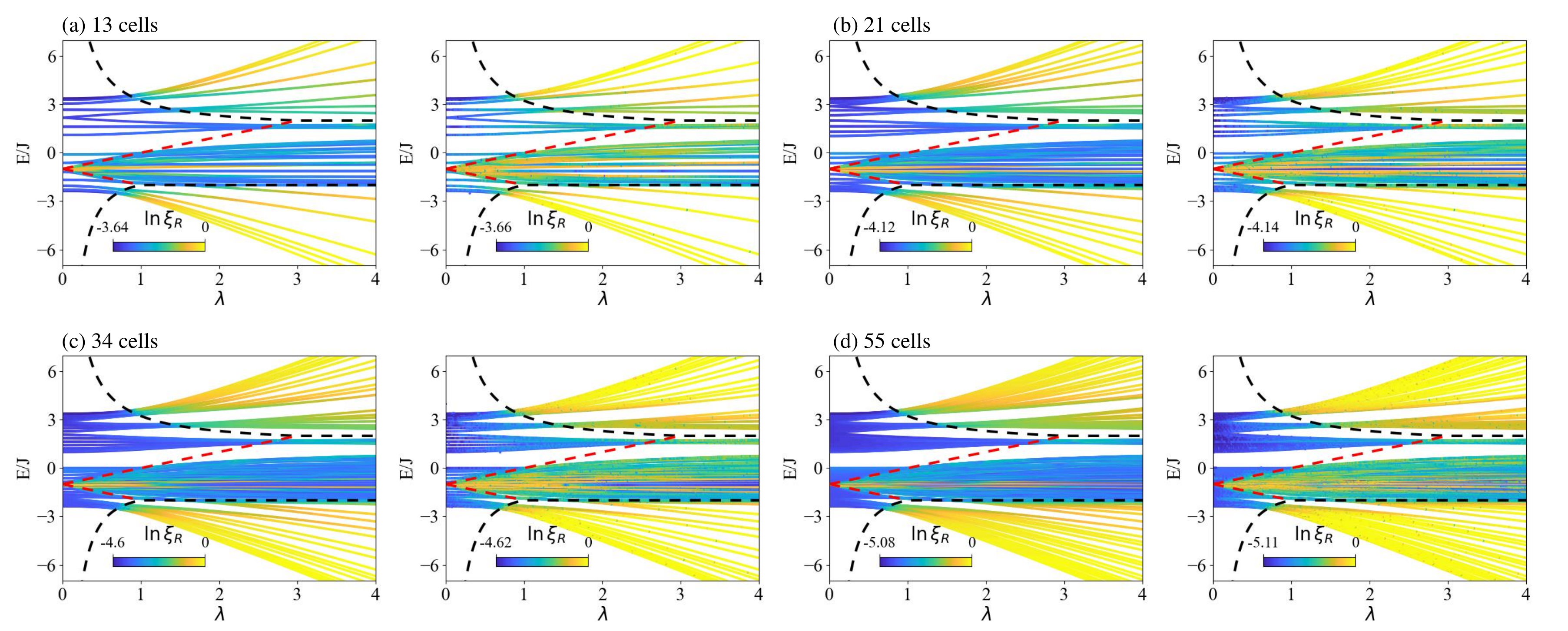}
\caption{Comparison of IPR between diagonalization calculation (the left column) and numerical simulated experiments (the right column) for the case of $13$ (a), $21$ (b), $34$ (c), and $55$ unit cells. Under the condition of the principal quantum number equals to $70$, the energy unit $J=2\pi\times7.53$MHz.}\label{ms4}
\end{figure*}

{Fig.~\ref{ms4} compares the IPR results obtained by diagonalizing the Hamiltonian~\eqref{eqdim} with those from the spectroscopicpic method of Ref.~\cite{PRoushan2017} under different unit cell sizes.}

{The results reveal that 13-unit-cell systems capture the universal critical behavior predicted by our theory, confirming experimental feasibility with existing Rydberg array technologies. As for the corresponding dual space IPR phase diagram, one can obtain them by conducting a Fourier transform on the wave functions in real space. Then, based on the dual space wave function, we can obtain the corresponding dual sapce IPR.}

\subsection{VII-3. Scaling of IPRs in real and dual spaces}

 \begin{figure}[htbp]
	\centering	
    \includegraphics[width=15cm]{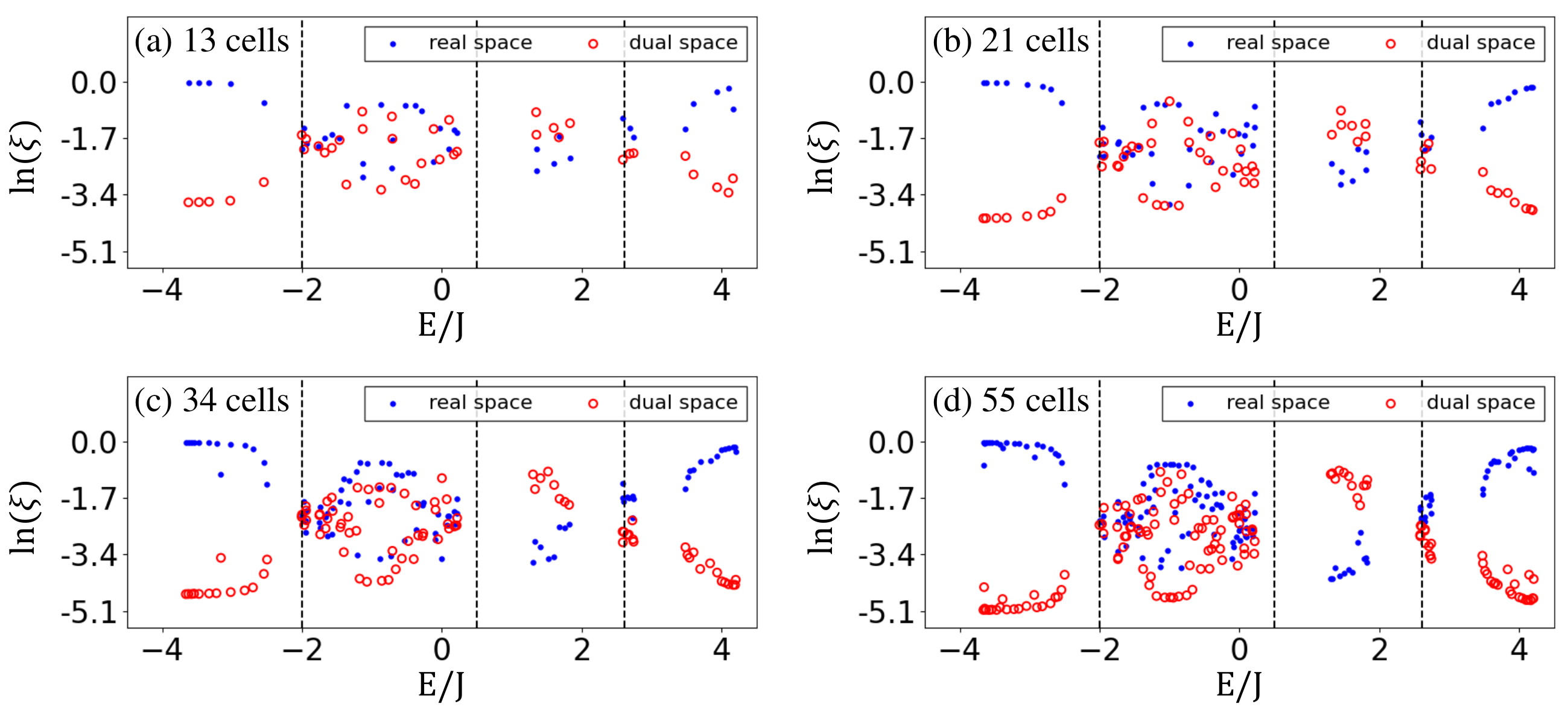}
	\caption{{The expected experimental IPRs for 13 (a), 21 (b), 34 (c) and 55 (d) cells. The lattice space IPR $\xi_R$ and the dual space IPR $\xi_K$ as functions of energy $E/J$}}\label{IPR_scaling}
\end{figure}

{ We exhibit the IPRs of $\lambda=1.5$ for $N=13,~21,~34,~55$ in the real space and dual space. As described in the main text, the key signatures are as follows: Localized/extended states: Real-space $\xi_R$ and dual-space $\xi_K$ IPRs exhibit spatial separation, with $\xi_R > \xi_K$ for localized states or $\xi_R < \xi_K$ for extended states. Multifractal states: $\xi_R$ and $\xi_K$ hybridize since $\xi_R \sim \xi_K$. One can find that $13$-unit-cell system can capture key properties of MMEs. }

\subsection{VII-4. The corresponding LEs}
{A localized state wave function satisfies the expression
\begin{equation}\label{fitting}
  \psi_{R/K}\propto \max\{\text{P}_{\beta,m}^{R/K}\}e^{-\gamma_{R/K}(m-m_0)},
\end{equation}
where $\max\{\text{P}_{\beta,m}^{R/K}\}$ is the maximum amplitude with a fixed $E_\beta$. $m_0$ is the site index of the location of maximum amplitude. LE ($\gamma_{R/K}$) is another quantity to characterize the three  phases and phase boundaries. From the results of $\text{P}_{\beta,m}^{R/K}$, one can directly obtain the corresponding eigenstates, and then by fitting the eigenstate exponentially, one can obtain the corresponding LEs}.
 \begin{figure}[htbp]
	\centering	
    \includegraphics[width=13cm]{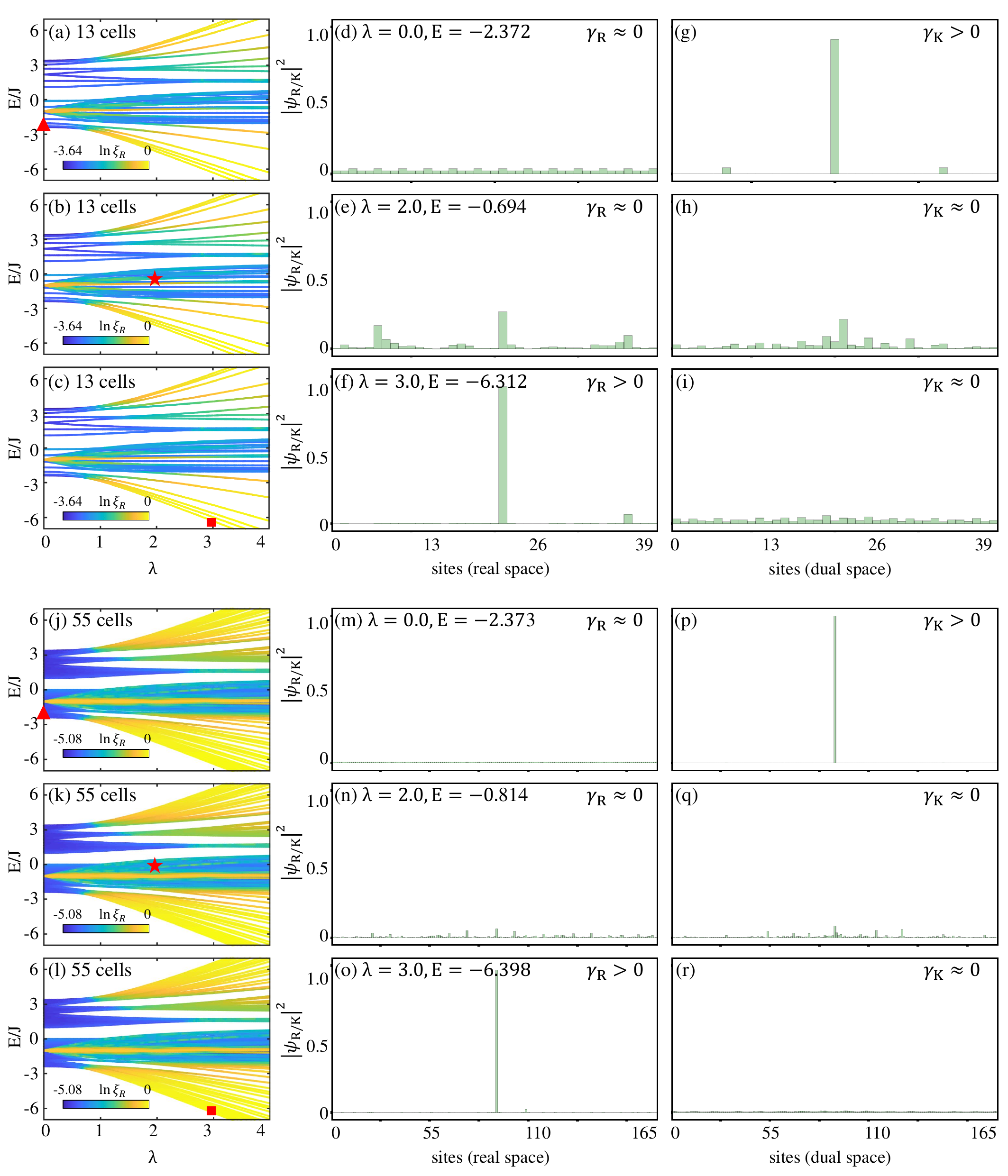}
	\caption{The expected experiments LEs with $13$ (a-i) and $55$ (j-r) unit cells. From top to bottom, the results correspond to the cases of extended, multifractal and localized states, respectively. Throughout, $J=2\pi\times7.53$MHz.}\label{LE54}
\end{figure}

{One can select an arbitrary point with fixed disorder strength $\lambda$ and energy parameters $E/J$ in the localized phase region, where the corresponding eigenstates will exhibit localization properties ($\gamma_{R}>0$) in real space and extension properties ($\gamma_{K}\approx 0$) in dual space. Similarly, wave function in the extended phase shows the extension property ($\gamma_{K}\approx 0$) in real space and the localization property in dual space ($\gamma_{K}>0$). Multifractal states are in between ($\gamma_{R}\approx 0$,~$\gamma_{K}\approx 0$).
We can perform exponential fitting on the eigenstate wave functions extracted through spectroscopic techniques, thereby obtaining the corresponding expected experiment LEs. Then, based on the above criteria, one can determine the properties of different phase regions. The typical data are provided(see Fig.~\ref{LE54}).}

{As shown in Fig.~\ref{LE54}, selecting three parameter points (triangle, star, square) in three phase regions, one can obtain wave functions of extended, multifractal and localized states, respectively. Through the characteristics of both real and dual-space LEs, one can effectively distinguish three different phases.}


\subsection{VII-5. Self-similarity of multifractal states}

\begin{figure}[h]
	\centering	
    \includegraphics[width=16cm]{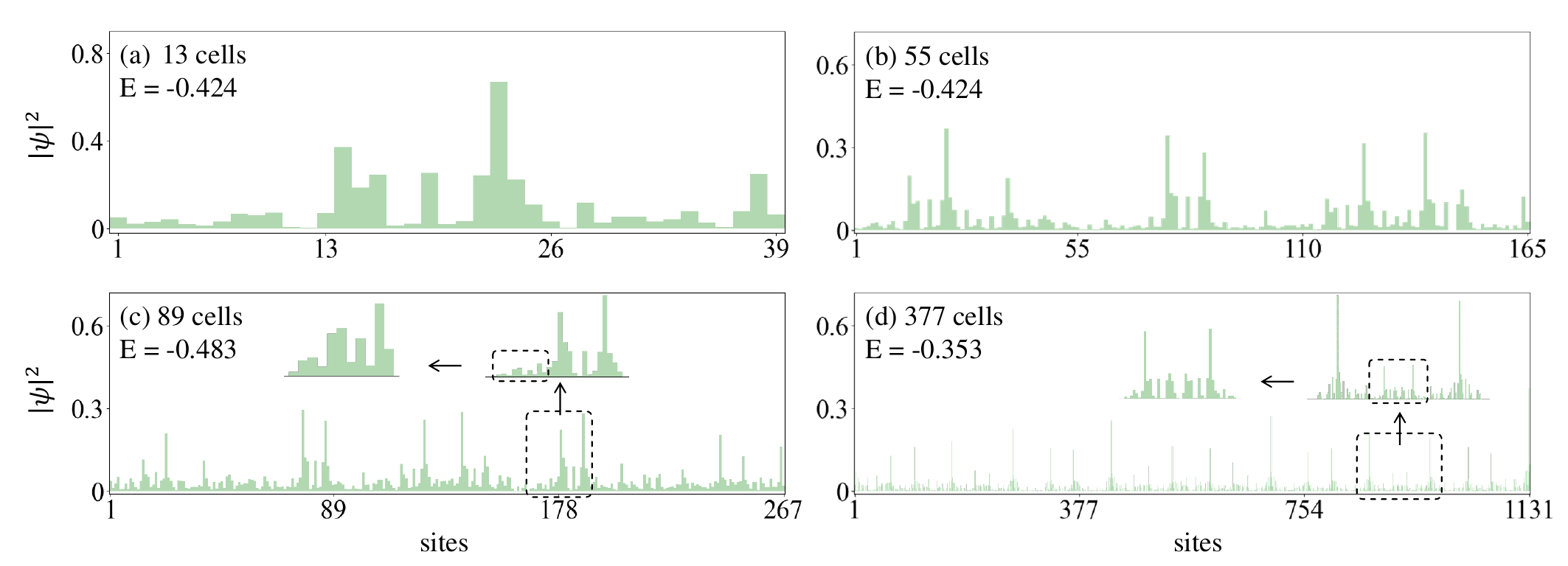}
	\caption{Self-similar structure of multifractal wave fucntions with a fixed $\lambda=2$ . The number of unit cells are marked. }\label{377}
\end{figure}

{The wave function with multifractal structure is self-similar. To demonstrate its self-similarity,  here in Figure~\ref{377}, we plot cases with the unit cell number $N=13$ (a), $N=55$(b), $N=89$ (c), and $N=377$(d). Self-similarity begins to emerge at system sizes N=13 and 55, and becomes much more pronounced at larger N values such as 89 and 377.}

{\emph{Summary.} We propose the experimental implementation and measurement schemes based on the Rydberg atomic array platform. Specifically, we demonstrate a spectroscopic technique capable of measuring IPRs across real-space and dual-space (see Fig.~\ref{ms4} and Fig.~\ref{IPR_scaling}). Based on this, we provide unambiguous criteria for determining different phase regions (Extended states:~$\xi_R<\xi_K$; Localized states:~$\xi_R>\xi_K$; Multifractal states: $\xi_R$ and $\xi_K$ hybridize since $\xi_R \sim \xi_K$), which are in good agreement with the phase boundary determined by the analytical solution. Furthermore, through this spectroscopic technique, we can also obtain the eigenstates' information. Then, LEs can be extracted through exponential fitting, so as to double check different phases and phase boundaries, i.e., Extended states:~$\gamma_R\approx0$ and $\gamma_K>0$; Localized states:~$\gamma_R>0$ and $\gamma_K\approx0$; Multifractal states: $\gamma_R\approx0$ and $\gamma_K\approx0$ (see Fig.\ref{LE54}).  Finally, although this manscript confirms that the MMEs and the three-state-coexisting quantum phase predicted by our theory can be experimentally verified through 13-unit-cell (tens of qubits), however, a larger size can no doubt better demonstrate the self-similarity of the multifractal wave function (see Fig.\ref{377}).}

\end{document}